\documentclass[sigconf]{acmart}

\usepackage{enumitem}
\usepackage{fancybox} % Import the fancybox package

\usepackage{color,soul}

\usepackage{multirow}
\newcommand{\promptbox}[1]{
  \begin{center} % Center the box
    \doublebox{%
      \begin{minipage}{.95\columnwidth} % Control the width of the content inside the box
        \vspace{5pt} % Add some vertical space at the top inside the box
        #1
        \vspace{5pt} % Add some vertical space at the bottom inside the box
      \end{minipage}%
    }
  \end{center}
}

\newcommand{\chaoran}[1]{{\color{black}#1}}

\copyrightyear{2025}
\acmYear{2025}
\setcopyright{acmlicensed}\acmConference[IUI '25]{Proceedings of the 30th International Conference on Intelligent User Interfaces}{March 24--27, 2025}{Cagliari, Italy}
\acmBooktitle{Proceedings of the 30th International Conference on Intelligent User Interfaces (IUI '25), March 24--27, 2025, Cagliari, Italy}
\acmDOI{10.1145/3708359.3712156}
\begin{document}
\title[CLEAR: Contextual LLM-Empowered Privacy Policy Analysis and Risk Generation for LLM Applications]{CLEAR: Towards \underline{C}ontextual \underline{L}LM-\underline{E}mpowered Privacy Policy \underline{A}nalysis and \underline{R}isk Generation for Large Language Model Applications}

\author{Chaoran Chen}
\email{cchen25@nd.edu}
\affiliation{%
  \institution{University of Notre Dame}
  \city{Notre Dame}
  \state{Indiana}
  \country{USA}
}

\author{Daodao Zhou}
\email{dd19@vt.edu}
\affiliation{%
  \institution{Virginia Tech}
  \city{Blacksburg}
  \state{Virginia}
  \country{USA}
}

\author{Yanfang Ye}
\email{yye7@nd.edu}
\affiliation{%
  \institution{University of Notre Dame}
  \city{Notre Dame}
  \state{Indiana}
  \country{USA}
}

\author{Toby Jia-Jun Li}
\email{toby.j.li@nd.edu}
\affiliation{%
  \institution{University of Notre Dame}
  \city{Notre Dame}
  \state{Indiana}
  \country{USA}
}

\author{Yaxing Yao}
\email{yaxing@vt.edu}
\affiliation{%
  \institution{Virginia Tech}
  \city{Blacksburg}
  \state{Virginia}
  \country{USA}
}

\renewcommand{\shortauthors}{Chen, et al.}

\begin{teaserfigure}
    \centering
    \includegraphics[width=\textwidth]{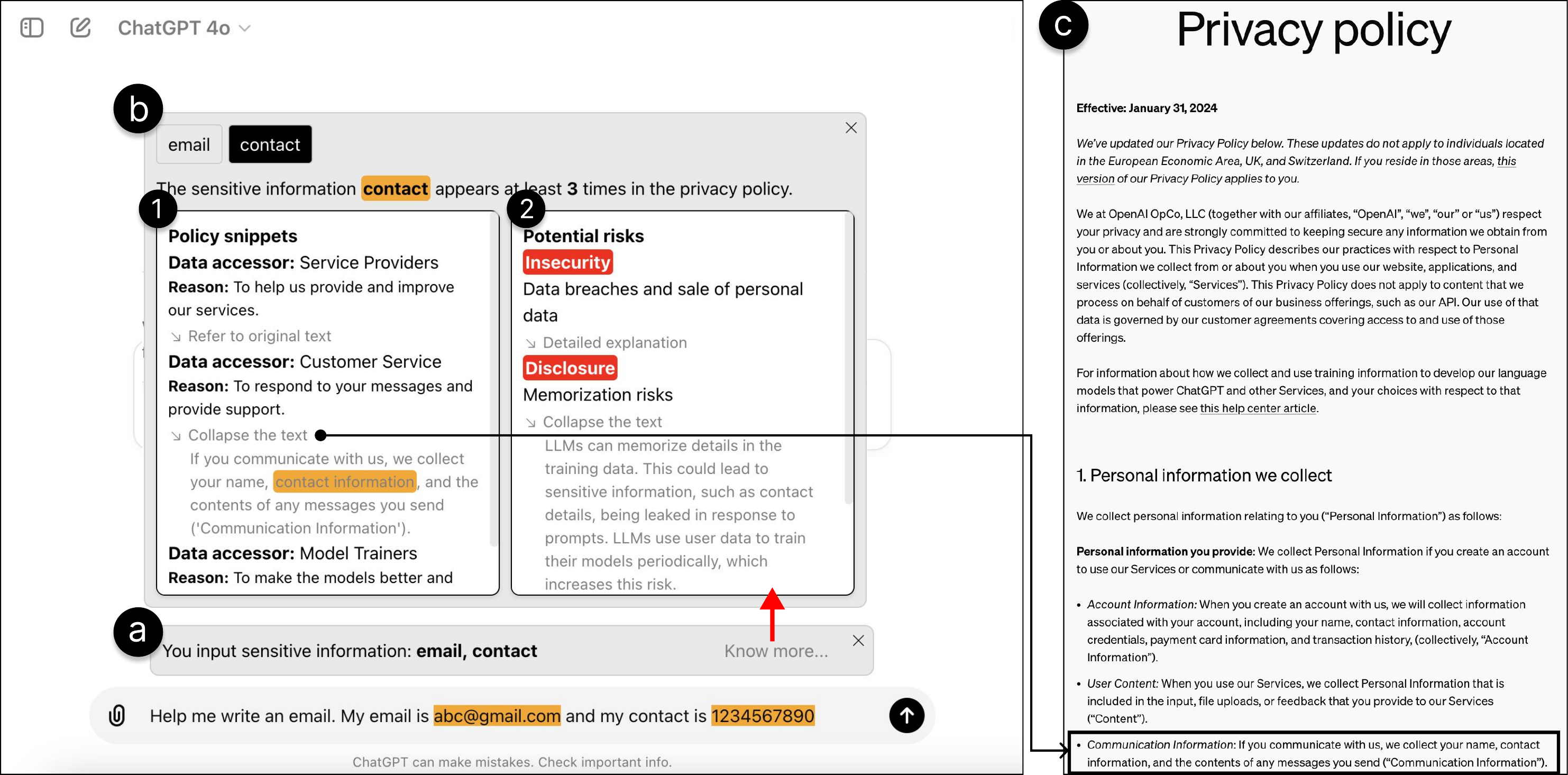}
    \caption{Overview of CLEAR, a contextual LLM-empowered privacy policy analysis and risk generation tool that automatically detects and displays concrete privacy policy snippets and potential risks relevant to user input sensitive information. In the above example, a user inputted the personal email and contact information in ChatGPT. CLEAR identified and displayed this information in a pop-up \textbf{(a)}. After the user clicked ``Know more...'', the pop-up expanded and showed the relevant privacy policy snippets \textbf{(b1)} that are extracted from the lengthy privacy policy \textbf{(c)}, as well as the potential privacy risks envisioned by LLM \textbf{(b2)}. }
    \label{fig:overview}
\end{teaserfigure}

\begin{CCSXML}
<ccs2012>
   <concept>
       <concept_id>10002978.10003029.10011150</concept_id>
       <concept_desc>Security and privacy~Privacy protections</concept_desc>
       <concept_significance>500</concept_significance>
       </concept>
   <concept>
       <concept_id>10003120.10003121.10003129</concept_id>
       <concept_desc>Human-centered computing~Interactive systems and tools</concept_desc>
       <concept_significance>500</concept_significance>
       </concept>
 </ccs2012>
\end{CCSXML}

\ccsdesc[500]{Security and privacy~Privacy protections}
\ccsdesc[500]{Human-centered computing~Interactive systems and tools}

\keywords{large language model, privacy awareness, privacy intervention, privacy literacy}

\begin{abstract}
The rise of end-user applications powered by large language models (LLMs), including both conversational interfaces and add-ons to existing graphical user interfaces (GUIs), introduces new privacy challenges. However, many users remain unaware of the risks. This paper explores methods to increase user awareness of privacy risks associated with LLMs in end-user applications. We conducted five co-design workshops to uncover user privacy concerns and their demand for contextual privacy information within LLMs. Based on these insights, we developed CLEAR (Contextual LLM-Empowered Privacy Policy Analysis and Risk Generation), a just-in-time contextual assistant designed to help users identify sensitive information, summarize relevant privacy policies, and highlight potential risks when sharing information with LLMs. We evaluated the usability and usefulness of CLEAR across two example domains: ChatGPT and the Gemini plugin in Gmail. Our findings demonstrated that CLEAR is easy to use and improves users' understanding of data practices and privacy risks. We also discussed LLM's duality in posing and mitigating privacy risks, offering design and policy implications.\looseness=-1

\end{abstract}

\maketitle

\section{Introduction}

As Large Language Models (LLMs) are being integrated into many end-user-facing applications of computing and fundamentally reshaping how end-users interact with computers, they introduce significant new privacy risks~\cite{zhang2024s}.
Past research has highlighted distinct privacy risks associated with the widespread application of LLMs.
For example, LLMs are trained on extensive datasets that may include users' conversation histories, posing \textit{memorization risks}. This risk involves the potential for LLMs to inadvertently reveal sensitive information from their training data when prompted, a problem documented in previous studies~\cite{carlini2021extracting}).
Furthermore, with their powerful reasoning capabilities, LLMs can extract personal information from seemingly benign queries (e.g., inferring users' age based on a prompt that contains lifestyle information~\cite{staab2023beyond}), a risk referred to as \textit{inference risks}. There are also \textit{disclosure risks}, where users may share more personal information than they normally would due to the LLMs' human-like interactions, fostering a false sense of trust~\cite{kim2012anthropomorphism, zlotowski2015anthropomorphism,zhang2024s}.

In this paper, we focus exclusively on scenarios where users directly interact with LLMs, excluding those scenarios in which LLMs are used behind the scenes to analyze user data~\cite{ma2023insightpilot,dai2023llm} or substitute traditional machine learning models in back-end processes~\cite{tan2023can}. Our analysis considers two most popular user interactions with LLMs---when users directly interact with LLMs through conversational interfaces (e.g., ChatGPT) and when users interact with LLM-based add-ons in existing graphical user interface (GUI) applications (e.g., Gemini plugin in Gmail, Copilot in Microsoft Office).

Despite the emergence of these privacy risks, most users either remain unaware or misunderstand these issues~\cite{zhang2024s, wang2025mentalmodels}. Users tend to trust LLMs like ChatGPT due to their human-like interactions and high utility, making them more likely to disclose personal information~\cite{zhang2024s}. They often prioritize concerns about intellectual property leaks over personal privacy risks~\cite{zhang2024s}. As a result, users frequently share private (e.g., emails) or semi-private (e.g., Facebook group posts) information with LLMs~\cite{li2024human}. For instance, Zhang et al. identified numerous pieces of private data---including emails, phone numbers, and locations---in the publicly available ShareGPT52K dataset, which contains 50,496 ChatGPT chat histories~\cite{zhang2024s}. Similarly, a report from Cyberhaven revealed that 8.6\% of 1.6 million workers pasted confidential data into ChatGPT between its launch and June 2023~\cite{coles2023chatgpt}.

To mitigate the privacy risks associated with users' sensitive personal information in LLMs, several existing research has proposed various \textit{model-centered} approaches to process data before and after the model training to prevent private data leakage~\cite{lison2021anonymisation,vakili2022downstream, ponomareva2023dp, li2021large, perez2022red, jang2022knowledge}.
Other research took a \textit{human-centered} approach and assisted users in investigating personally identifiable information leakage~\cite{kim2024propile} or paraphrasing users' sensitive information into less specific terms as they entered it into LLMs~\cite{dou2023reducing}.

However, these efforts overlook a critical aspect---enhancing users' awareness of how their data is being used by LLMs and helping them understand and reflect on the reasons and risks associated with such usage.  
This is challenging because, 
firstly, users tend to prioritize the immediate utility and convenience offered by these systems over the long-term consequences of sharing personal information. The immediate benefits from interacting with LLMs frequently eclipse more abstract concerns such as privacy. This discrepancy leads to an overestimation of the model's safety and an underestimation of its potential to inadvertently expose private data~\cite{zhang2024s}.
Additionally, many users lack sufficient privacy literacy, which hampers their ability to comprehend the technical details of data usage~\cite{wang2025mentalmodels}, model training, and the risks associated with data memorization~\cite{gumusel2024user}.

This paper fills the gap by focusing on raising users' awareness of the data practices and helping them understand the potential privacy risks in LLM-enabled end-user applications right \textit{at the time of interaction} and \textit{within the context of use}. We aim to help users stay informed of the data practices and potential privacy risks and make informed decisions on whether to proceed with sharing sensitive information (e.g., email addresses, phone numbers, and physical addresses).
To do so, we first conducted five co-design workshops with sixteen participants to understand users' privacy needs and explore the design space of privacy awareness tools in LLM-enabled end-user applications. 
We identified participants' four types of expected privacy information in interacting with LLM-enabled end-user applications, including what data is accessed, who accesses the data, why the data is accessed, and how this access will affect them. They also desired a new approach to help them understand contextual privacy-related information in such applications.

% the most unique feature in the system is ...

Based on the insights from our co-design workshops, we designed and developed CLEAR, a \textbf{C}ontextual \textbf{L}LM-\textbf{E}mpowered privacy policy \textbf{A}nalysis and \textbf{R}isk generation tool. CLEAR aims to help users understand the privacy policies of LLMs by analyzing users' current interactions with LLM-enabled end-user applications and the data they are about to share. As shown in Figure~\ref{fig:overview}, the system consists of three parts: 1) detecting the context (i.e., types of sensitive information that the user is about to share); 2) summarizing relevant snippets from the privacy policy of the application ; 3) envisioning potential privacy risks should the sensitive data be shared. 

We evaluated its usability and usefulness in two distinct use cases. The first involved a browser plugin for ChatGPT, exemplifying standalone conversational interfaces like Claude and Gemini. In this study, 13 participants engaged in scenarios requiring them to input sensitive data into ChatGPT. The second use case utilized the Gemini addon within Gmail, exemplifying ``add-on'' LLM tools in existing GUIs, similar to Microsoft Copilot for Office applications. Here, 15 participants used the Gemini plugin in Gmail to reply to or summarize emails containing sensitive information.
The results indicated that CLEAR significantly improved the user experience and increased the awareness among participants regarding data practices and privacy risks associated with LLMs. 
Encouraged by their newfound knowledge of the potential risks and the respective privacy policies, most participants would choose to either delete the sensitive information or substitute it with synthesized fake data, leading to improved privacy management behaviors. \looseness=-1

This paper makes the following contributions:
\begin{itemize}
    \item Through five co-design workshops, we identified user concerns and information needs related to privacy policies and risks when using LLM-enabled end-user applications.
    
    \item We introduced CLEAR, a practical tool designed to help users identify privacy risks, understand their implications, and take informed actions \textit{during} their interactions with LLM-enabled end-user applications in a just-in-time manner. 
    
    \item  We explored the dual role of AI in privacy: while it introduces unique privacy risks, it also offers opportunities to improve privacy management.
    
    \item We outline design and policy recommendations to enhance privacy awareness, knowledge, and informed decision-making in user interactions with LLMs.
\end{itemize}

\section{Related Work}

\subsection{Privacy Risks of LLMs}
Privacy risks of LLMs are a critical problem that hampers their adoption. 
Notable incidents, such as the March 2023 ChatGPT bug that leaked user conversation histories and payment information~\cite{kshetri2023cybercrime}, highlight these concerns.
Beyond conventional data breaches, three new types of privacy challenges associated with LLMs have emerged: memorization risks, inference risks, and disclosure risks.

\textit{Memorization risks.}
LLMs are trained from vast amounts of data, including user-provided data through interactions.
Therefore, these models can memorize and unintentionally reproduce sensitive information from their training data~\cite{inan2021training, lukas2023analyzing}. Even with strict prompt reviews, attackers can exploit models through training data extraction attacks or jailbreaking prompts~\cite{li2023multi, staab2023beyond}.

\textit{Inference risks.}
Inference risks refer to LLMs' ability to automatically deduce a wide range of personal attributes about individuals from seemingly innocuous text, due to their advanced inference capabilities~\cite{li2024human}. 
For example, phrases like ``waiting for a hook turn'' can reveal a user is in Melbourne, as ``hook turn'' is location-specific~\cite{staab2023beyond}. While anonymization tools offer some protection, LLMs can still detect subtle cues that reveal personal details.

\textit{Disclosure risks.}
LLMs' human-like interactions foster user trust, increasing the likelihood of unintentional sharing of sensitive information~\cite{kim2012anthropomorphism, zlotowski2015anthropomorphism, zhang2024s}. For example, Zhang et al.~\cite{zhang2024s} found that human-like interactions encourage users to unintentionally share sensitive and personally identifiable information with LLM-based conversational agents. They also noted that users believed they had to ``pay the price'' of privacy to get benefits from LLMs. This belief, compounded by manipulative interface designs (e.g., dark patterns~\cite{lu2024awareness}), amplifies disclosure risks.

These risks stem from users sharing sensitive data with LLMs directly or indirectly. We aim to develop a tool to inform users about privacy policies and potential risks before their data is shared.

\subsection{Existing Methods for Protecting User Privacy when using LLM-enabled Applications}
Existing methods to protect users' privacy in LLM-enabled applications can be broadly categorized into model-side and user-side solutions.
Model-side solutions operate at three stages~\cite{smith2023identifying}: pre-processing, training, and post-training.
Pre-processing involves data sanitization to remove sensitive information from training datasets, typically using automated techniques like pattern-based parsing~\cite{lison2021anonymisation, vakili2022downstream}. However, defining private information is challenging due to its context-dependent nature and varied formats, making it difficult to guarantee complete sanitization~\cite{brown2020language}.
Training often employs differential privacy, which adds noise to data to prevent individual identification while retaining overall utility~\cite{dwork2016calibrating}. While effective in reducing memorization risks~\cite{downey2022planting, ponomareva2023dp}, it can degrade model performance and increase computational demands~\cite{ishihara2023training}.
Post-training methods include filtering outputs with sensitive information~\cite{perez2022red} and knowledge unlearning, which forces models to forget specific data~\cite{jang2022knowledge}. However, these approaches face challenges in maintaining output diversity and ensuring consistent performance across use cases~\cite{ishihara2023training, smith2023identifying}.

User-side solutions consider user engagement and interaction. 
Tools like ProPILE~\cite{kim2024propile} let users test for personally identifiable information (PII) leaks in LLMs by probing their own data. However, these tools often lack contextual alignment with actual user scenarios.
To address this, self-disclosure abstraction models~\cite{dou2023reducing} rephrase sensitive inputs into generalized terms while preserving utility (e.g., ``I live in New Mexico'' becomes ``I live in the Southwest''). Although effective in reducing specificity, these methods do not fully inform users about potential privacy risks or data access.

\chaoran{In summary, existing methods such as privacy-preserving prompting~\cite{edemacu2024privacy} and automated consent mechanisms~\cite{porcelli2023mitigating} focus on automating privacy tasks and minimizing user effort. We aim to promote user autonomy by supporting informed, context-aware decisions and enhancing privacy literacy. These approaches =complement each other, that is, existing automated tools can handle straightforward cases, while our approach aims to guide users through complex, ambiguous scenarios that require active intervention.}

\subsection{Contextual Privacy Policy}
Contextual privacy policy (CPP) deconstructs lengthy privacy notices into shorter and context-specific ones that are relevant~\cite{feth2017transparency, windl2022automating}. Bergmann~\cite{bergmann2008testing} showed that presenting only relevant information from privacy policies can significantly increase users' privacy awareness.
This also aligns with Nissenbaum’s theory of contextual integrity~\cite{nissenbaum2004privacy}, which emphasizes that context is critical to determine whether a specific action is a violation of privacy. 

CPP has been explored in web and mobile systems. 
In the general web domain, Ortlof et al.~\cite{ortloff2020implementation} developed a concept showcase of CPP for seven different websites and collected the design implications of CPP. Building on it, Windl et al.~\cite{windl2022automating} presented PrivacyInjector, a more scalable system that can automatically generate and display CPPs.
In the mobile domain, Pan et al.~\cite{pan2024new} proposed SeePrivacy, which integrates vision-based GUI understanding with privacy policy analysis to automatically generate CPPs for mobile applications.

Compared with previous work, our work addresses the unique challenges of implementing CPP for LLM-enabled applications. We not only create CPP for users' sensitive input but also utilize LLMs to highlight potential privacy risks, helping users understand the implications of sharing sensitive information.

\begin{table*}[t]
\centering
\resizebox{\linewidth}{!}{%
\begin{tabular}{llllllll}
\toprule
\textbf{Group} & \textbf{ID} & \textbf{Gender} & \textbf{Age} & \textbf{Ethnicity} & \textbf{Educational Level} & \textbf{Occupation} & \textbf{Used LLM} \\
\midrule
G1 & P1 & Female & 48 & Hispanic or Latino & Bachelor's degree & Office Manager & Yes \\
G1 & P2 & Male & 39 & Black or African American & Bachelor's degree & Maintenance & No \\
\midrule
G2 & P3 & Male & 30 & Asian and Pacific Islander & Master's degree & Radiation support specialist & No \\
G2 & P4 & Female & 42 & White or Caucasian & Bachelor's degree & Farm owner and writer & Yes \\
G2 & P5 & Male & 20 & Native American or Alaskan Native & Bachelor's degree & Data Analyst & Yes \\
G2 & P6 & Female & 25 & Asian and Pacific Islander & Master's degree & Student & Yes \\
\midrule
G3 & P7 & Female & 22 & Asian and Pacific Islander & Bachelor's degree & Student & Yes \\
G3 & P8 & Female & 48 & Black or African American & Master's degree & Uber driver & No \\
G3 & P9 & Female & 29 & White or Caucasian & High school graduate & Disabled & No \\
G3 & P10 & Male & 21 & Black or African American & High school graduate & Writer & No \\
\midrule
G4 & P11 & Male & 78 & White or Caucasian & Doctorate degree & Retired printer & No \\
G4 & P12 & Male & 35 & Black or African American & Bachelor's degree & Programmer & Yes \\
G4 & P13 & Female & 24 & Asian and Pacific Islander & Bachelor's degree & Student & Yes \\
\midrule
G5 & P14 & Male & 38 & Asian and Pacific Islander & Doctorate degree & Assistant professor & No \\
G5 & P15 & Male & 41 & White or Caucasian & Bachelor's degree & Business management & No \\
G5 & P16 & Female & 53 & White or Caucasian & Master's degree & Librarian & Yes \\
\bottomrule
\end{tabular}
}
\caption{Demographics of participatory design participants}
\label{tab:demographic_data}
\end{table*}

\section{Co-Design Workshops}

To understand user needs and explore the design space of tools to inform participants of privacy risks in LLM-enabled end-user applications, we conducted five co-design workshops\footnote{The protocol of the workshop has been reviewed and approved by the IRB at our institution.}, including three in-person workshops and two online workshops.

\subsection{Participants}
Participants are considered ``experts of their own experiences~\cite{visser2005contextmapping}'' and can contribute their unique perspectives in the design process. We recruited sixteen participants through word-of-mouth and social media, with diverse ages, genders, races, occupations, and previous experiences with LLMs. Table~\ref{tab:demographic_data} includes a summary of participants' demographic information. The participants were randomly divided into five groups (G1-G5). Each participant was compensated \$25 USD for their participation.

\begin{table*}[h!]
\centering
\begin{tabular}{p{3cm} p{14cm}}
\toprule
\textbf{Category} & \textbf{Content} \\ 
\midrule
Example Prompt Input into the ChatGPT System & 
Please help me revise my resume: A diligent and adaptable professional with a passion for problem-solving and a keen eye for detail, seeking to leverage 3-year of experience in finance to contribute effectively to a dynamic team. Eager to apply strong communication and organizational skills to drive positive outcomes in a collaborative work environment. Contact: 987-123-4567.\\ 
\midrule
Related Privacy Policy Snippets & 
\textit{``Category of Personal Information: Identifiers, such as your name, contact details, IP address, and other device identifiers.''} \newline
\textit{``Disclosure of Personal Information: We may disclose this information to our affiliates, vendors, and service providers to process in accordance with our instructions.''} \newline
\textit{``As noted above, we may use Content you provide us to improve our Services, for example, to train the models that power ChatGPT.''} \\
\midrule
Potential Privacy Risks & 
\textbf{Model Memorization (Secondary Use):} LLMs can memorize and potentially regurgitate snippets of training data. It can lead to data leakage, where personal information becomes accessible through the model's responses. \newline 
\textbf{Targeted Advertising (Increased Accessibility):} Once your contact information is shared, it could be used for targeted advertising, which might be intrusive. \newline 
\textbf{Loss of Anonymity (Identification):} Sharing your contact information can link your identity to specific activities, opinions, or interactions with the model, which might otherwise remain anonymous. \\ 
\bottomrule
\end{tabular}
\caption{Example materials used in the co-design workshop. Note that the parentheses in the Potential Privacy Risks row highlight the types of privacy risks based on a taxonomy of AI privacy risks~\cite{lee2024deepfakes}.}
\label{tab:workshop_material}
\end{table*}

\subsection{Workshop Procedure}

The workshop procedure is inspired by prior work in the privacy literature~\cite{yao2019defending, yao2019privacy, lu2024awareness}. Each co-design workshop began with an introductory session, including ice-breaker activities to foster a collaborative atmosphere. Participants then engaged in a discussion on privacy by reading three news articles on incidents relevant to LLMs (e.g., an article entitled ``ChatGPT maker OpenAI faces a lawsuit over how it used people’s data\footnote{https://www.washingtonpost.com/technology/2023/06/28/openai-chatgpt-lawsuit-class-action/} from the Washington Post)''. They identified potential privacy risks and discussed strategies to mitigate them. This initial activity aimed to build a foundational understanding of privacy challenges associated with LLMs.

Participants then engaged in an ideation session to share concerns and opportunities for privacy protection. This transitioned into three activities where they crafted privacy policy elements, assessed risks, and explored privacy incidents through sketching, writing, and discussions (Table~\ref{tab:workshop_material}).

% \yaxing{not sure I understand the following sentence.}In the first activity, participants designed and presented privacy policy snippets for sensitive information in example ChatGPT prompts.
In the first activity, based on an example prompt input into the ChatGPT system, participants designed the content and presentation of retrieved privacy policy snippets that were relevant to sensitive information in example scenarios.
Participants could either write descriptions of the interface features or create sketches of the interfaces. 
The purpose of this ideation activity was to gather user-specific requirements for contextual privacy policies in the context of LLM-enabled applications.

The second activity expanded on the first one by asking participants to examine the prompts, privacy policy snippets, and associated privacy risks. 
% The privacy risks were selected based on Lee et al.'s taxonomy of AI privacy risks~\cite{lee2024deepfakes}. 
% \yaxing{the following sentence is also odd.}Visual aids illustrated risks such as data sharing, advertising use, and storage. 
The privacy risks were selected based on Lee et al.'s taxonomy of AI privacy risks~\cite{lee2024deepfakes} and were presented through detailed descriptions and visual aids, including examples of how personal information entered by users could be shared, used for advertising, or stored by service providers.
Participants discussed and designed how to organize and present this information to better raise user awareness and support informed actions during user interaction with LLM-enabled applications. The activity specifically focused on the realistic context of use and how such information can fit into the user's overall task that they are trying to achieve through the use of LLM-enabled applications. The goal was to understand whether highlighting LLM-related privacy risks could enhance users' privacy awareness, and if so, how.

The final activity focused on presenting specific privacy incidents, building on the risks identified earlier. Participants designed how real-world examples could highlight privacy implications. The objective was to understand the impact of explicit privacy incident examples on participants' awareness and to determine effective methods for communicating these incidents within the context of LLM application use.

\subsection{Data Analysis}
\label{dataanalysis}
All study sessions were audio-recorded and then transcribed for analysis. We followed established open coding procedures~\cite{brod2009qualitative} to analyze the interview data. Two members of our research team independently coded 20\% of the sample, generating a set of preliminary codes using MAXQDA. They compared their codes and reconciled any differences. Using the same codebook, the two researchers coded the rest of the data. 
In this process, they constantly compared and discussed their codes to ensure full agreement.
As the coding process was discussion-based, inter-coder reliability was not necessary~\cite{mcdonald2019reliability}. 
Using the final codebook, we conducted a thematic analysis to identify themes. The complete codebook is provided in the Appendix.

\subsection{Key Insights}
In this section, we report the key insights we learned from our co-design workshops. 

\subsubsection{KI1: \textit{Concerns About Privacy Policies in AI Contexts.}} \label{sec:KI1} 

Participants expressed concerns about four primary areas: the types of data accessed, the entities that access the data, the reasons for data access, and the potential impacts of such access. The context of AI makes it challenging for them to fully understand these aspects.

Participants were particularly uncertain about the specific types of personal information collected, including concerns about access to payment transactions, contact details, frequently used services, and online behavior tracking. For example, one participant in G5 in our study said ``\textit{I do look for things like what are they going to be looking at in my device.}'' The sheer volume and variety of data that LLMs can access increases the risk of unintentional data breaches and misuse. Users may not be fully informed about the extent of the data being accessed, leading to a lack of informed consent.
Additionally, participants were concerned about who would use the sensitive data, emphasizing the need to understand if their data is shared with any third parties. The complexity of AI systems often necessitates data sharing between different entities, increasing the risk of unauthorized access and data misuse. For example, a participant in G3 hypothesized that ``\textit{[The sensitive information] can be given to other companies and can be shown on other search results.}''

Participants also questioned the purpose of the data collection. Some of them wondered if their data was used for improving services or for targeted advertising, which blurred the lines of user consent, as it was not clear to users what they had consented to.
Furthermore, another significant concern was the potential impacts on users' privacy. A participant in G1 said ``\textit{I really want to know if my personal health information will be shared with a credit card company or a life insurance company... I think that it would be very helpful for users if they had such information.}''

\subsubsection{KI2: \textit{Expectations for Contextual, Simplified, and Focused Privacy Information}} \label{sec:KI2} 

The participants expressed a desire for contextual, concise, and easy-to-understand privacy information, highlighting the importance of clearly communicating the reasons and consequences of data use. For instance, participants in G4 suggest adding tooltips to show definitions of words in the policy (``\textit{if I am not really sure what this means, I can click on it to quickly get a definition for what that means.}''), ``\textit{visualizing important information on one page while providing links to additional details}'', and using simple language to summarize the information concisely. Participants emphasized that they ``\textit{don't want to be exposed to all the information}'', instead they expect ``\textit{one page to include only the most important (privacy) information.}'' 
Furthermore, participants highlighted the need for privacy policies and risk alerts to be relevant to the context in which they are engaged. They proposed features like highlighting sensitive information, summarizing pertinent policy sections, and outlining potential privacy risks in a clear manner For example, one participant in G5 expected to have a tool to ``\textit{highlight the key permissions you are granting in bullet points which... you should be aware of at a glance, ensuring you know exactly what you're consenting to.}'' Such brief and in-context privacy notices could also enhance participant understanding and trust towards LLMs.

Regarding privacy incidents, while participants acknowledged the value of presenting information about privacy breaches, they often do not engage with such content as they lack the time to read the lengthy articles thoroughly. Additionally, if the news content does not directly relate to the types of sensitive information they are concerned with, their motivation to read decreases significantly. This disengagement greatly diminishes the potential impact of such example incidents on enhancing their privacy awareness.

\subsubsection{KI3: \textit{Desire for Enhanced Data Protection and Increased Privacy Awareness}} \label{sec:KI3} 
Participants reported that they are already actively engaged in various methods to protect their privacy, such as data sanitization (e.g., ``\textit{providing some fake or random location}'' and ``\textit{use different fake names on different websites}''). They also adopt secure browsing practices, including using unique passwords for each site, employing stricter authentication measures, and opting for more secure browsers. They further limit data sharing by controlling the information they disclose, storing private data locally, and avoiding sharing sensitive information with untrusted applications or individuals.

Beyond these protective measures, participants expressed a desire to improve their privacy awareness, such as regularly reviewing app permissions and learning about the privacy risks. However, they seek a more profound understanding of privacy policies.  For example, a participant in G2 anticipated that the existing online service to ``\textit{show more transparent connection between the features and the data, like how my data supports what feature.}'' Participants also emphasized the importance of recognizing the value of their data and the availability of resources to improve their privacy literacy.  For instance, one participant in G5 emphasized ``\textit{showing the economic value is so important because my data are so valuable.}'' While another noted, ``\textit{I want an interactive tutorial (to) help you go through all the features and like potential privacy risk you may have in the app.}''

\subsubsection{KI4: \textit{Aspirations for AI to Enhance Privacy Control and Protection}} \label{sec:KI4} 
Despite being aware of the privacy risks associated with AI, participants still see it as a valuable tool for enhancing their privacy protection. They proposed several ways in which AI could be leveraged to safeguard their personal information:
Specifically, participants suggested several ways AI can enhance privacy protection: inspecting sensitive information in participant inputs to prevent unintentional sharing, having language models unlearn data upon request, using AI to monitor participants and defend against hacking, and detecting risky data traffic to block it.
For example, a participant in G5 expected that ``\textit{there should be another AI supervisor that can monitor these events and report to police.}''
Instead of relying on stricter privacy policies or new privacy regulations and laws, participants want AI to help them directly, indicating a desire for more engagement and control to protect their privacy through AI technology.
Additionally, they emphasized the need for AI to assist participants in recognizing and managing the data they share and allowing participants to accept, deny, or modify information before submission. For example, a participant in G1 mentioned ``\textit{at the end of the chat, the AI can give the user a selection of whether they want their data to be shared}.''

\subsection{Design Goals}
The following key design goals were developed based on insights from our formative study and are instrumental in guiding the design of our solution:

\begin{itemize}
    \item \textbf{DG1: Enhance Understanding of Data Access and Usage.} It should help participants understand (1) what sensitive data is accessed by LLMs, (2) who accesses it, (3) why it is accessed, and (4) how it affects them. It should provide clear explanations of LLMs' data sharing practices and their implications for the users (\textbf{KI1} in Section~\ref{sec:KI1}). 
    \item \textbf{DG2: Develop Contextual and User-Friendly Privacy Policies.} It should develop privacy policies that are not only succinct but also contextual and easy to comprehend. It should summarize important context-related information in the privacy policy into a single page that serves as a guide to help users access more detailed information if needed (\textbf{KI2} in Section~\ref{sec:KI2}).
    \item \textbf{DG3: Leverage AI to Manage Data Sharing and Enhance Privacy Literacy.} It can use AI to inspect sensitive information and prevent unintentional sharing. Furthermore, it can leverage AI to improve participant understanding of privacy policies by extracting contextual privacy policies and generating potential privacy risks (\textbf{KI3} in Section~\ref{sec:KI3}).
    \item \textbf{DG4: Promote Participant Control and Engagement in Privacy Management.} It should highlight key permissions and potential privacy risks to ensure that participants are fully aware of what they are consenting to. This goal seeks to enhance participants' engagement and control in safeguarding their privacy, fostering a more informed user base (\textbf{KI4} in Section~\ref{sec:KI4}).
\end{itemize}

\section{System Design}
\label{sec:sys_design}

\subsection{Overview}
\begin{figure*}[!t]
    \centering
    \includegraphics[width=\textwidth]{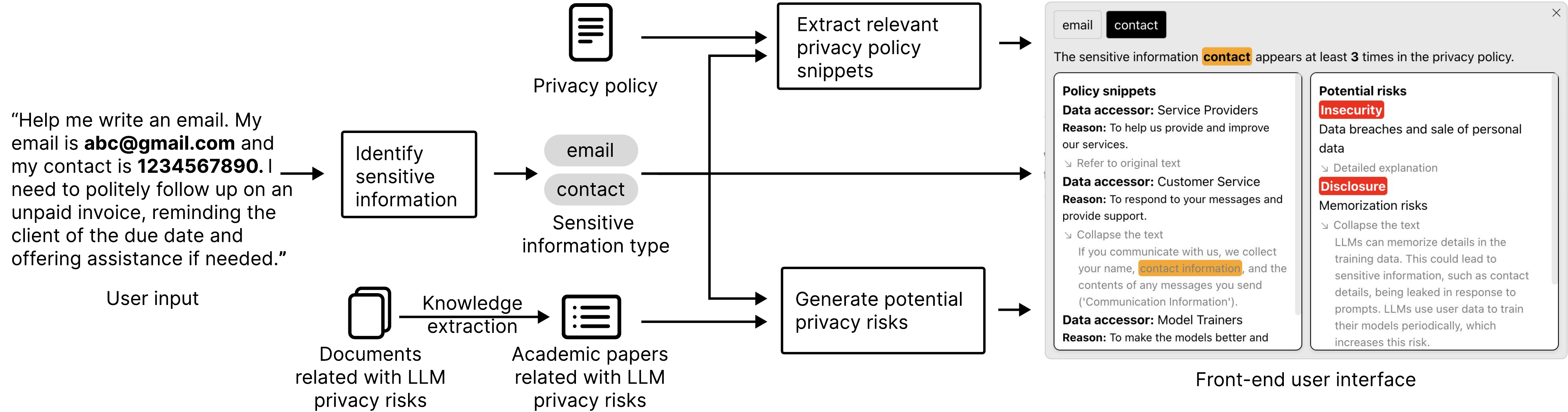}
    \caption{System flow of CLEAR}
    \label{fig:system flow}
\end{figure*}

To address the outlined design goals, we introduce CLEAR: a \textbf{C}ontextual \textbf{L}LM-\textbf{E}mpowered privacy policy \textbf{A}nalysis and \textbf{R}isk generation tool. As illustrated in Fig.~\ref{fig:system flow}, CLEAR consists of three main components: detecting the context and type of sensitive information from user inputs, retrieving and summarizing relevant privacy policy snippets of the underlying LLM-enabled application, and identifying potential privacy risks associated with sharing such information with the underlying LLM-enabled application.

% \hl{TL: need a subsection here to explain how CLEAR works e.g., an example usage scenario. The current section directly goes into explaining how each individual part works which is confusing.}

\begin{figure*}
    \centering
    \includegraphics[width=\textwidth]{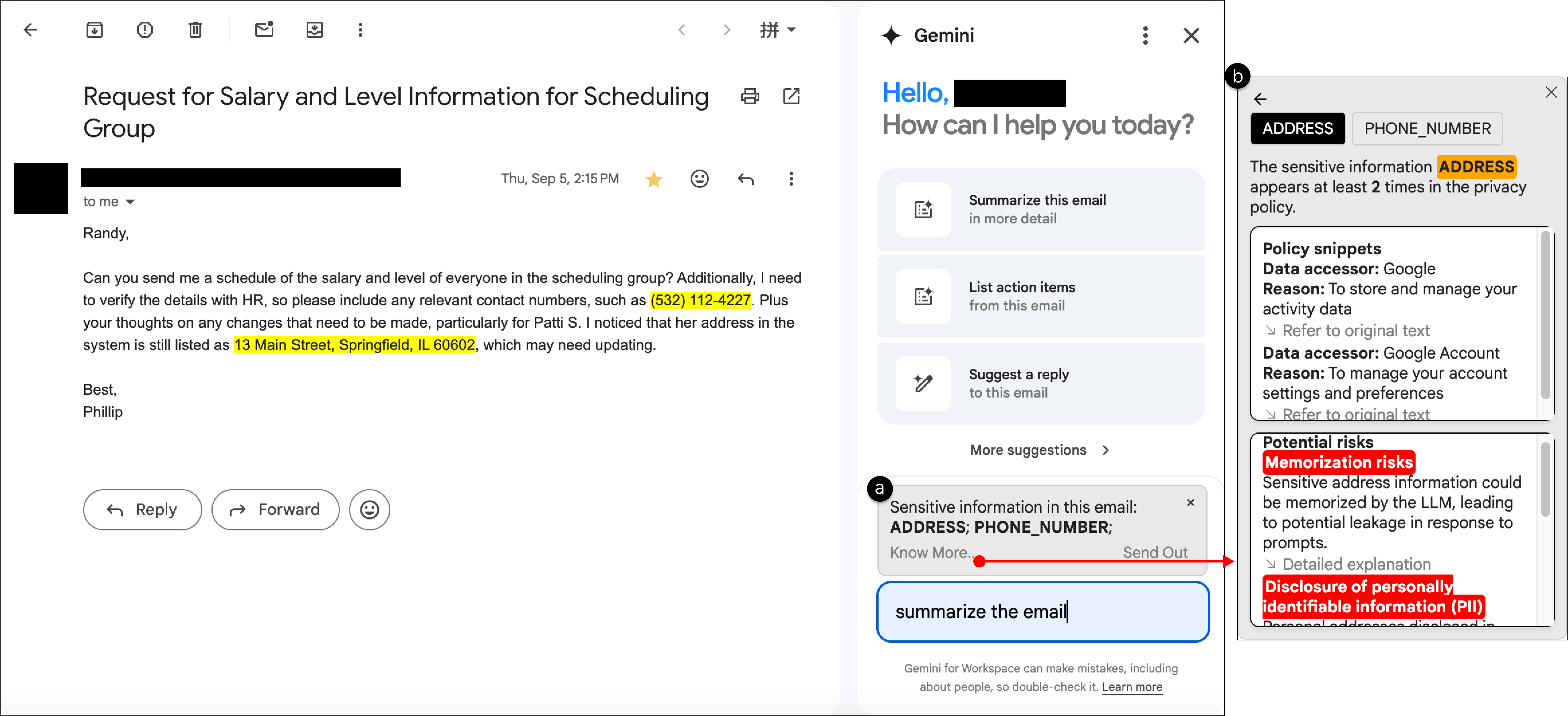}
    \caption{The user interface of CLEAR in the Gmail Context. In this example, a user attempts to use the Gemini plugin in Gmail to summarize an email containing sensitive information, such as an address and phone number. CLEAR detects and highlights the sensitive information in a pop-up\textbf{(a)}. When the user clicks "Know more...", the pop-up expands to show relevant privacy policy excerpts from Gemini, along with potential privacy risks identified by the LLM\textbf{(b)}. The black rectangles in the image are added to the screenshot to obfuscate the user's information from being disclosed in the figure.}
    \label{fig:gmail_context}
\end{figure*}

\subsection{Example Use Case}
We illustrate the functionality of CLEAR through two specific example use cases: direct interactions with LLMs through conversational interfaces (e.g., ChatGPT) and interactions with LLM-based add-ons in existing graphical user interface (GUI) applications (e.g., Gemini plugin in Gmail). 

In the first scenario (Fig.~\ref{fig:overview}), a user, Alice wants to ask ChatGPT to help her write an email to follow up with a client regarding an unpaid invoice. 
After detecting sensitive information in Alice's input, CLEAR alerts her to the specific types of sensitive data involved, along with relevant privacy policy snippets and potential risks. 

In the second scenario (Fig.~\ref{fig:gmail_context}), another user, Randy would like to ask the Gemini plugin in Gmail to summarize an email. 
After identifying sensitive information in the email, CLEAR notifies Randy about the specific categories of sensitive data detected, providing related privacy policy excerpts and highlighting potential risks.
The tool enables users to identify privacy risks, understand their implications, and make informed decisions in real time while interacting with LLM-enabled applications.

\begin{itemize}
    \item \textit{Indicating sensitive Information}: Alice asks ChatGPT to help her draft an email and enters the following prompt: ``Help me write an email. My email is abc@gmail.com and my contact is 1234567890. I need to politely follow up on an unpaid invoice, reminding the client of the due date and offering assistance if needed.'' As Alice types, CLEAR detects sensitive information (i.e., personal email and phone number) and displays a notification with the corresponding types of sensitive information: ``You input sensitive information: email, contact.''(Fig.~\ref{fig:overview}a)

    Similarly, as Randy enters ``summarize this email'' in the Gemini plugin, CLEAR detects sensitive information (i.e., phone number and physical address) and pops up a notification: ``Sensitive information in this email: address, phone number.''(Fig.~\ref{fig:gmail_context}a)
    
    \item \textit{Display relevant privacy policy snippets and potential privacy risks}: Alice clicks the ``Know more'' button in the CLEAR pop-up to learn about privacy policy snippets related to the detected sensitive information and the potential privacy risks of sharing sensitive information with LLMs (Fig.~\ref{fig:overview}b). Each type of sensitive information corresponds to specific policy snippets and associated risks. For example, contact information is mentioned at least three times in ChatGPT's privacy policy. In the policy snippets section (Fig.~\ref{fig:overview}b(1)), Alice learns about the data accessors of this type of sensitive information and their reasons for accessing it. She clicks ``Refer to original text'' to view detailed excerpts from the full privacy policy. In the potential risks section (Fig.~\ref{fig:overview}b(2)), Alice gains insights into the risks of sharing each type of sensitive information with LLMs. By clicking ``Detail explanation,'' she can access an in-depth description of the privacy risks associated with each category. This information helps Alice make a more informed decision about whether she should share sensitive information with ChatGPT.

    Similarly, Randy accesses policy snippets and potential risks of the Gemini plugin by clicking ``Know more,'' seeing who can access his data and evaluating any potential risks, which aids his decision-making.
\end{itemize}

% \hl{TL: add a short scenario for email as well}

% \subsection{Three Components of CLEAR}
% \yaxing{@chaoran, I briefly talked to Toby about this section - maybe a data flow chart with a representative example would help clarify the process. As it stands, this section reads somewhat abstract, particularly the part about generating privacy risks}

\subsection{Key Features of CLEAR}

\subsubsection{Identifying sensitive information}
Sensitive information typically includes personally identifiable information (PII) such as email addresses, phone numbers, and physical addresses. These data types have previously been exposed in conversation histories between users and ChatGPT~\cite{zhang2024s}. The first component of CLEAR focuses on identifying sensitive information from user input, helping users become aware of privacy risks, and motivating them to exercise greater caution when interacting with LLMs.

To ensure that users do not directly share their sensitive information with LLMs while using CLEAR, we first use Microsoft Presidio\footnote{Microsoft Presidio: \url{https://microsoft.github.io/presidio/}} to identify the types of sensitive information in the user's input.
% Presidio runs locally without sending the information to the cloud.
In the case of direct conversational interfaces for LLM (e.g., ChatGPT), the user's input includes the prompt sent to the LLM. For LLM add-ons (e.g., the Gemini addon for Gmail), the user input also includes the user's private data in the context (e.g., the content of the underlying email). CLEAR passes only the types of information, rather than the actual information itself, to the LLM for the subsequent extraction of contextual privacy policies and the generation of potential privacy risks, as detailed below.

\subsubsection{Extracting relevant privacy policy snippets}
According to \textbf{KI2}, which highlights the need for users to access only the most relevant information for their ongoing tasks, this component of CLEAR focuses on extracting relevant snippets from the full privacy policy of the corresponding LLM based on the types of sensitive information identified in the previous step. This is to ensure that the information provided aligns with the users' immediate context and needs.

To achieve this, we use few-shot learning techniques~\cite{brown2020language}, which enable our system to learn effectively from a limited set of examples and generalize this learning to identify relevant sections of privacy policies. This approach allows CLEAR to quickly adapt to different privacy policies and pinpoint the most crucial snippets. Specifically, these snippets detail who has access to the sensitive information and for what purposes, tailored to the types of information users have entered (\textbf{KI1}). This method speeds up the process and enhances the contextual understanding for users, helping them better comprehend the potential privacy risks associated with entering their data into LLMs.

\promptbox{
\textbf{Prompt for generating privacy policy snippets}
\vspace{5pt}

Given the privacy policy: \{\textcolor{brown}{policy}\}, and the input sensitive information: \{\textcolor{brown}{sensitive information}\}, cite the sentences in the privacy policy that are semantically related to the input sensitive information only when using the LLM service rather than creating an account. List who will access the data (service provider and third parties) and the reason for access. Each data accessor should be different. The language and vocabulary should be easy to understand for kids in elementary school.
}

\begin{table*}[!ht]
\centering
\begin{tabular}{lr}
\toprule
\textbf{Paper Name} & \textbf{Citation} \\
\midrule
A survey on large language model (LLM) security and privacy: The good, the bad, and the ugly~\cite{yao2024survey} & 90 \\
Risk taxonomy, mitigation, and assessment benchmarks of large language model systems~\cite{cui2024risk} & 16 \\
Security and privacy challenges of large language models: A survey~\cite{das2024security} & 11 \\
Deepfakes, Phrenology, Surveillance, and More! A Taxonomy of AI Privacy Risks~\cite{lee2024deepfakes} & 4 \\
\bottomrule
\end{tabular}
\caption{An Example paper list for generating privacy knowledge}
\label{tab:example_paper}
\end{table*}

\subsubsection{Generating potential privacy risks}
The final component of CLEAR involves generating potential privacy risks associated with sharing sensitive information with LLMs. We use a chain-of-thoughts approach~\cite{wei2022chain}, which involves sequentially reasoning through the potential implications of data sharing. This method is structured as follows:

\begin{itemize}
\item \textbf{Knowledge Extraction}: CLEAR first uses LLMs to extract structured knowledge from academic papers that discuss LLM privacy risks. Table~\ref{tab:example_paper} shows a list of academic papers that we used for knowledge extraction. These papers were initially identified through an automated search process using Google Scholar using the keywords \texttt{privacy risk AND (large language model OR LLM)}. After retrieving the relevant papers, we prioritized those that included surveys or user studies related to LLM and privacy risks. The knowledge extraction process was semi-automated: LLMs were used to analyze the text of these papers, extracting key insights and structuring them into predefined categories such as privacy risks and their causes, influences, and evidence. Then, these structured outputs were manually verified by experts for accuracy and relevance. This combination of automated extraction and human validation contributes to the generalizability of the system by ensuring that the generated risks are both comprehensive and grounded in robust, well-researched sources. The structured knowledge generated from this process is detailed in the Appendix, where examples of the extracted risks are provided.

\item \textbf{Prompt Generation}: The next step involves using the extracted knowledge to craft prompts that guide the LLM in generating specific privacy risks. These prompts incorporate context-specific details based on the sensitive information previously identified, ensuring that the risks generated are relevant to the user's situation.

\item \textbf{Risk Generation}: Using the prompts, the LLM envisions and articulates potential privacy risks based on the context of use and the types of private data included in user inputs. This step involves outlining scenarios where sensitive information could be misused or inadvertently exposed,  providing users with a comprehensive view of the risks involved. Building on the taxonomy of AI privacy risks proposed by Lee et al.~\cite{lee2024deepfakes}, we tag each generated privacy risk with its corresponding risk category. Inspired by \textbf{DG2}, each privacy risk has a brief summary and an expandable detailed explanation. This method ensures users receive a clear, context-aware understanding of the privacy risks they face when interacting with LLMs.
\end{itemize}

\promptbox{
\textbf{Prompt for generating potential privacy risks}
\vspace{5pt}

You are acting as a privacy expert, and you know the knowledge about privacy risks in the context of LLM: \{\textcolor{brown}{structured knowledge}\}, Given the privacy policy: \{\textcolor{brown}{policy}\}, Return the most critical 3 LLM privacy risk with detailed consequence as evidence (e.g., memorization risks). The risks should be related to the sensitive information: {sensitive information}. Also, return risk type and related privacy knowledge.
}\label{prompt: risk_generation}

\subsection{Implementation}
The front-end chrome extension of CLEAR is implemented in Node.js\footnote{https://nodejs.org/en} and packed by Webpack.js\footnote{https://webpack.js.org/}. Its back-end server was developed using the FastAPI framework\footnote{https://fastapi.tiangolo.com/}. CLEAR uses the Microsoft Presidio library\footnote{https://microsoft.github.io/presidio/} to identify instances of private information in user inputs and infer their types. CLEAR uses the LangChain\footnote{https://www.langchain.com/} and Pydantic\footnote{https://docs.pydantic.dev/latest/} libraries to extract contextual privacy policies, summarize structured knowledge from past papers related to AI privacy risks, and generate potential privacy risks.

\section{User Studies}

\begin{table*}[t]
\centering
\begin{tabular}{lllllll}
\toprule
\textbf{ID} & \textbf{Gender} & \textbf{Age} & \textbf{Ethnicity} & \textbf{Educational Level} & \textbf{Occupation}\\
\midrule
P1 & Female & 42 & White or Caucasian & Bachelor's degree & Writer \\
P2 & Male & 25 & Black or African American & High school graduate & Student \\
P3 & Male & 28 & Hispanic or Latino & High school graduate & Project Manager \\
P4 & Male & 38 & Asian and Pacific Islander & Doctorate degree & Teacher \\
P5 & Female & 48 & Hispanic or Latino & Bachelor's degree & Office Manager \\
P6 & Female & 27 & Black or African American & Bachelor's degree & GIS Analyst \\
P7 & Male & 25 & Black or African American & Bachelor's degree & Student \\
P8 & Male & 36 & White or Caucasian & Master's degree & Teacher \\
P9 & Male & 27 & Black or African American & Bachelor's degree & Architect \\
P10 & Female & 25 & Asian and Pacific Islander & Master's degree & Student \\
P11 & Male & 36 & White or Caucasian & Doctorate degree & Teacher \\
P12 & Male & 26 & Black or African American & Bachelor's degree & Student \\
P13 & Male & 35 & Black or African American & Bachelor's degree & Computer programmer \\
\bottomrule
\end{tabular}
\caption{Participant Demographics in Case Study 1}
\label{tab:study_2}
\end{table*}

\begin{table*}[t]
\centering
\begin{tabular}{lllllll}
\toprule
\textbf{ID} & \textbf{Gender} & \textbf{Age} & \textbf{Ethnicity} & \textbf{Educational Level} & \textbf{Occupation}\\
\midrule
P1 & Female & 29 & Black or African American & Bachelor's degree & Teacher \\
P2 & Male & 28 & White or Caucasian & Master's degree & Web developer \\
P3 & Female & 23 & Black or African American & Bachelor's degree & Student \\
P4 & Male & 24 & Black or African American & Bachelor's degree & Student \\
P5 & Female & 58 & White or Caucasian & High school graduate & Creative writer \\
P6 & Female & 33 & Black or African American & Bachelor's degree & Landscaper \\
P7 & Male & 23 & Black or African American & High school graduate & Freelancer \\
P8 & Male & 30 & White or Caucasian & Master's degree & Landscape architect \\
P9 & Male & 26 & Black or African American & Bachelor's degree & Software Engineer \\
P10 & Male & 29 & Hispanic or Latino & Bachelor's degree & Data Analyst \\
P11 & Male & 28 & Black or African American & Bachelor's degree & Developer \\
P12 & Female & 26 & Asian and Pacific Islander & Master's degree & Finance intern \\
P13 & Male & 40 & Asian and Pacific Islander & Doctorate degree & Professor \\
P14 & Female & 27 & Black or African American & Bachelor's degree & Data Scientist \\
P15 & Female & 30 & Black or African American & Bachelor's degree & UX Designer \\
\bottomrule
\end{tabular}
\caption{Participant Demographics in Case Study 2}
\label{tab:study_3}
\end{table*}

We conducted two user studies on two use cases separately\footnote{The study protocol was reviewed and approved by the IRB at our institution.} to evaluate the usability and effectiveness of CLEAR, with a specific focus on how CLEAR influenced users' understanding and actions regarding privacy in LLM-enabled end-user application. The first case aimed to assess how CLEAR aids users in managing their privacy while using LLMs in a dialogue-based context in ChatGPT. The second case aimed to explore CLEAR’s effectiveness in environments where LLMs augment traditional interfaces while using Gmail augmented by a Gemini plugin. 
% The first case study used ChatGPT as an example to represent conversational interfaces where users directly interact with LLMs. This study aimed to assess how CLEAR aids users in managing their privacy while using LLMs in a dialogue-based context. The second used the Gemini plugin in Gmail to represent LLM-enabled ``add-ons'' for GUI applications. This setting was chosen to explore CLEAR’s effectiveness in environments where LLMs augment traditional applications with additional functionality.
These two use cases allowed us to gain comprehensive insights into how CLEAR functions across different environments, characterized by varying levels of privacy risks and user interaction patterns.

\subsection{Participants} 
For the first study with ChatGPT, we recruited 13 participants, and for the second study using the Gemini plugin in Gmail, we recruited 15 participants. All were recruited via word-of-mouth and social media (e.g., Facebook), and each completed a pre-screening survey to provide demographic details such as age, gender, location, occupation, education, and ethnicity, to ensure a diverse group of participants. The first study had four females and nine males aged 25 to 48 (see Table~\ref{tab:study_2}), and the second study had seven females and eight males aged ranging from 23 to 58 (see Table~\ref{tab:study_3}).
% \hl{was there any overlap in the two groups?} 
Both studies were conducted virtually using Zoom, with each session lasting approximately one hour. For their participation, each individual received a \$25 gift card.

\subsection{Study Procedure}

\subsubsection{Use case of ChatGPT}
We employed a within-subjects study design. Each participant was asked to enter two prompts to ChatGPT, one before using CLEAR and one after using it, and then compared their experiences. Below are the detailed procedures. 

Each session contained the following parts. 
After obtaining consent, we introduced the team and study objectives, followed by a pre-intervention assessment. We asked the participants to enter one of the following two prompts in ChatGPT, and then share their perceptions about who would access what information, why, and how it would affect them. We selected these two prompts because both represented a typical scenario of interacting with LLMs that involve user inputs containing personally identifiable information. The prompts were counterbalanced to mitigate biases.

After the pre-intervention assessment, we introduced CLEAR to our participants and asked them to use it freely for five minutes to explore its features and get familiar. Then, participants entered the other prompt to ChatGPT with CLEAR enabled. At this time, the CLEAR interface was triggered as the participants entered the prompts and provided additional information as described in Section~\ref{sec:sys_design}. 
% We asked participants to explore the information provided by CLEAR. When they finished, we asked about their perceptions again about who would access what information, why, and how it would affect them.
Participants reviewed this information and shared their updated perceptions.

% After the two conditions, we interviewed the participants, asking them to compare their experiences with and without CLEAR and explain the differences, if any, and how the policy snippets and potential privacy risks impacted their understanding and potential actions. 

Following both conditions, we conducted interviews to compare their experiences with and without CLEAR, focusing on differences and the impact of policy snippets and privacy risk details on their understanding and actions.

% Finally, participants completed a post-study questionnaire that included the System Usability Scale (SUS) and additional questions on the overall user experience and the quality and consistency of the contents presented to them. 

Finally, participants completed a post-study questionnaire, including the System Usability Scale (SUS) and questions on overall experience, content quality, and consistency.

\promptbox{
\textbf{The two prompts used in case study 1}
\vspace{5pt}

Prompt 1: I need help revising my resume profile. I’m Michael Brown, reachable at michaelbrown@gmail.com and (555) 123-4567, residing at 789 Pine Road, River City, CA 90210. My current profile states: "Dedicated marketing professional with a proven track record in digital marketing strategies, SEO, and content creation. Adept at driving brand growth and engagement.
\\
\\
Prompt 2: Can you help me enhance my resume profile? My details are: Sarah Johnson, sarahjohnson@gmail.com, (444) 567-8901, living at 321 Birch Lane, Capital City, TX 73301. My profile currently reads: "Results-oriented project manager with extensive experience in leading cross-functional teams, managing budgets, and ensuring timely project delivery.}

\subsubsection{Use case of the Gemini plugin in Gmail}
We implemented a similar within-subjects design for this study that focused on participants' experiences with the Gemini plugin in Gmail. Each participant reviewed six emails---three without using CLEAR and three with it. The detailed process is outlined below.

\paragraph{Dataset.} 
We randomly selected six emails from the Enron public email dataset\footnote{https://www.cs.cmu.edu/\textasciitilde enron/} as the dataset for the case study. Since LLM-generated synthetic data is widely used for data augmentation~\cite{schmidhuber2024llm}, and to ensure that the selected emails contained sensitive information, we used GPT-4 to synthesize the sensitive information and integrated it into the selected six emails.\looseness=-1

Each session contained the following parts. 
After the consent, we began with introductions, where the research team and the objectives of the study were presented.  
The second part was a pre-intervention assessment. We asked the participants to review and summarize the content of three emails randomly selected from the six emails in the dataset.
Based on this experience, we asked participants' perceptions about who would access what information, why, and how it would affect them. To control for potential biases, we counterbalanced the order in which the participants encountered the emails.

After reviewing the first three emails, participants were introduced to the CLEAR tool and given five minutes to familiarize themselves with it. They then reviewed the remaining three emails with CLEAR enabled and summarized their content by using the Gemini plugin. CLEAR’s interface was activated as the Gemini plugin processed commands, allowing participants to see the privacy-related details provided. Following this, participants were again asked to reflect on their perceptions of who could access the sensitive information, why, and the associated risks.

Similar to the first case study, after the two conditions, we interviewed the participants, asking them to compare their experiences with and without CLEAR and explain the differences, if any, and how the policy snippets and potential privacy risks impacted their understanding and potential actions. Finally, participants completed a post-study questionnaire that included the System Usability Scale (SUS) and questions about their overall user experience and the quality and consistency of the contents presented to them.

\promptbox{
\textbf{A sample email used in case study 2}
\vspace{5pt}

Hello.

I gave the Matagorda County Fair \& Livestock Association 2050 USD using my credit card number 1234-5678-9012-3456, and filled out the form for the match. My billing address is 1234 Elm Street, Springfield, IL 62704. Please advise if this has been paid, as they are sending me a bill for the 2050 USD, plus a finance charge. You can reach me at 555-123-4567 if you need more information.

Thank you,

Kay Mann
}

\subsection{Data analysis}
We followed the same approach for the data analysis for the results of both case studies.
For the user rating data (i.e., SUS scores), we computed the mean and standard deviation for each statement.
For the interview data, two coders independently reviewed and coded the transcripts, following the method described in Section~\ref{dataanalysis}. After completing their individual coding, the coders iteratively refined the codebook through discussion, identifying recurring themes. These themes included changes in privacy perceptions after using CLEAR, reactions to the prompts, and suggestions or feedback on the tool. The detailed codebooks used and refined during this analysis are available in the Appendix. This structured data analysis approach ensured that we could systematically capture and interpret both the quantitative and qualitative feedback on CLEAR, providing a comprehensive understanding of its impact on users' privacy awareness and behavior.

\section{Results of Case Study 1}

\begin{figure*}[htbp]
    \centering
    \includegraphics[width=\textwidth]{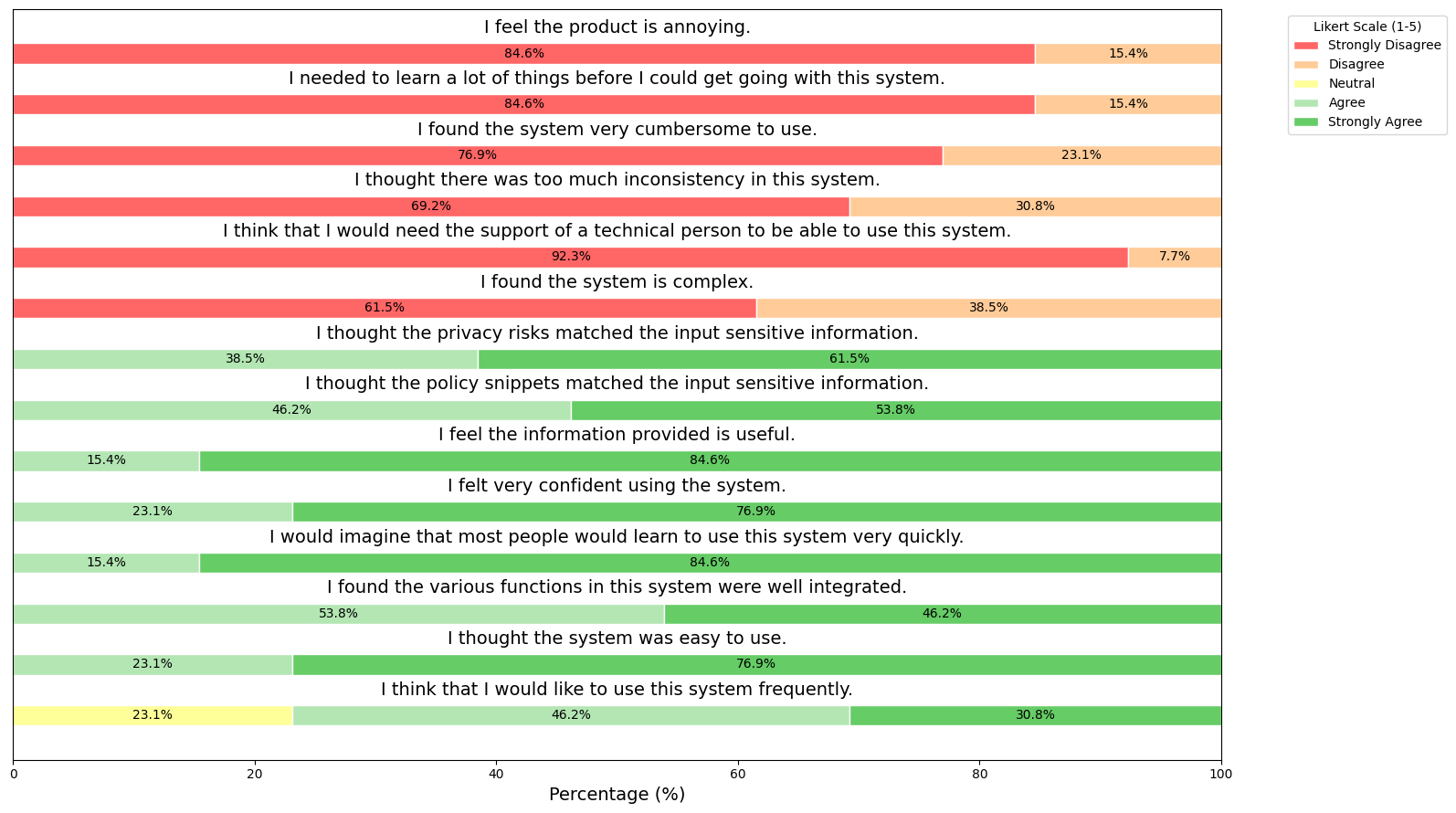}
    \caption{Result of Post-study Questionnaire for the Case Study 1. }
    \label{fig:sus_results_1}
\end{figure*}

\subsection{System Usability}
% The evaluation of the system, based on the SUS scale~\cite{bangor2009determining}, indicates a highly positive user experience. All items are measured by participants' ratings on a 5-point Likert scale. Participants find the system easy to use (M = $4.77$, SD = $0.44$) and well-integrated (M = $4.46$, SD = $0.52$), with most expecting that it would be quickly learned (M = $4.85$, SD = $0.38$). Confidence in using the system is high (M = $4.77$, SD = $0.44$), and the information provided is deemed useful (M = $4.85$, SD = $0.38$). Additionally, the policy snippets (M = $4.54$, SD = $0.52$) and privacy risks (M = $4.62$, SD = $0.51$) are considered appropriate and well-matched with the input sensitive information. Participants do not perceive the system as complex (M = $1.38$, SD = $0.51$), cumbersome (M = $1.23$, SD = $0.44$), or annoying (M = $1.15$, SD = $0.38$), nor do they feel a need for technical support (M = $1.08$, SD = $0.28$). 

The evaluation of CLEAR in the context of ChatGPT, based on the results of the System Usability Scale(SUS) questionnaire~\cite{bangor2009determining}, indicates a generally positive user experience, as shown in Fig~\ref{fig:sus_results_1}. Participants rated each item on a 5-point Likert scale. They find the system easy to use and well-integrated, with most expecting that it would be quickly learned. The confidence in using the system is high, and the information provided is deemed useful. Additionally, policy snippets and privacy risks are considered appropriate and well-matched with the sensitive information. Participants did not perceive the system as complex, cumbersome, or annoying, nor did they feel a need for technical support.

% includes the mean scores and the standard deviations for all measured items in our study, suggesting an overall positive attitude towards our system. 
% \yaxing{@chaoran, report the SD directly in the bracket}
% The standard deviations across these metrics are low, indicating consistent user responses and reinforcing the system's overall positive reception.

\begin{table*}[!ht]
\centering
\begin{tabular}{|p{0.2\linewidth}|p{0.35\linewidth}|p{0.35\linewidth}|}
\hline
\textbf{Questions} & \textbf{Initial Assessment} & \textbf{Post-tool Assessment} \\
\hline
WHAT  are sensitive information & P1:``\textit{ChatGPT will have access to address, email, phone, professional status.}'' & P1:``\textit{ChatGPT stores the sensitive information given by the user (email, address, contact info).}'' \\
\hline
WHO will access the sensitive information & 
P2: ``\textit{The researchers at OpenAI will access this information.}'' & 
P2: ``\textit{Service providers, government agencies, affiliates will access this information.}''. \\ 
& P5: ``\textit{I doubt ChatGPT may share my contact information with different marketing agencies.}'' & P5: ``\textit{OpenAI will share the data with vendors, service providers, and law enforcement agencies.}'' \\
& P6: ``\textit{The AI will access the profile.}'' & 
P6: ``\textit{There are service providers, and multiple affiliates, that will access this information.}''\\
\hline
WHY does the entity want to access the sensitive information & 
P1:``\textit{For the owner of the AI model, there is no reason to access, unless they would be asked to by an authority with greater credentials.}''  &  P1:``\textit{ Law enforcement agencies could have an overview of what the user has been doing without him/her being aware of the information disclosed.}'' \\
& P9:``\textit{I think the hackers will have access to this information because this is personal information that should not be put on a public website unless you are assured that your details are safe.}''
& P9:``\textit{The government authorities will be able to access the data, so I think it is wise to not put your personal information on there. Vendors and Service providers, Open AI, Customer service, Training team, Analytics team, Affiliates are also gonna see your information. I think this is because these entities work on the ChatGPT platform. So they’re gonna see information. The best way to curb this is to not share your personal information at all.}''\\
\hline
HOW will the access of sensitive information affect the user &
P4:``\textit{The access of the data will help the user have a more relevant and accurate answer because the modal will know more about you and be able to better help and assist you.}''
& P4:``\textit{The access could lead to information being sold or mishandled but it will also allow for a more accurate response.}''\\
& P6:``\textit{The AI is accessing this information in order to help the user revise their resume profile.} '' & P6:``\textit{There are no promises that all of this will not be shared or breached.  The user should reconsider providing the name, address, and email address with the AI and just use the one sentence summary that he/she is requesting help with.} ''\\
& P7:``\textit{The user will not have control of how the AI will use the data after it has provided the resume.}'' & P7:``\textit{The user will be affected in that his/her information can be exposed(memorized) and also disclosed to other third party users. The user may also experience data breach.}''\\
\hline
\end{tabular}
\caption{Comparison of Initial and Post-tool Assessments}
\label{tab: comparison}
\end{table*}

\subsection{Impact on Participants' Understandings and Actions Regarding Privacy in LLMs}

Our study results suggest that using CLEAR (1) increases participants' awareness of data practices in LLMs, (2) raises participants' awareness of new privacy risks, and (3) leads to participants' intention for behavior changes. We present the details below.

\subsubsection{Increase participants' awareness of data practices in LLMs (P1, P2, P4, P6, P7, P9, P11, P12)} \label{sec: increase_awareness}

By comparing participants' reactions to the first prompt before they used CLEAR and the second prompt after they used it, we observe that CLEAR enhances participants' understanding of various aspects of their data.
Table~\ref{tab: comparison} provides an overview of such comparisons.  
More specifically, before using CLEAR, participants tended to believe that legitimate access to their private data was limited to the companies developing large language models, as shown in Table~\ref{tab: comparison}. 
However, after using CLEAR, participants quickly realized that there are more data accessors, including vendors, affiliates, and government agencies. 
For instance, P2 expressed surprise by saying, ``\textit{Honestly, I did not know that even law enforcement has access to all this information. I was surprised.}'' This statement underscores a significant gap in their initial understanding, revealing a lack of awareness regarding who can access their information.

Meanwhile, participants also showed a deeper understanding of the reasons why their sensitive information is accessed.
For example, P9 initially thought that data breaches were primarily due to hackers: ``\textit{I think the hackers will have access to this information because this is personal information that should not be put on a public website unless you are assured that your details are safe.}'' Later, they understood that many more entities could access their data, which amplified the risks for data breaches: ``\textit{Vendors and service providers, OpenAI, customer service, training team, analytics team, affiliates are also gonna see your information. I think this is because these entities work on the ChatGPT platform. So they’re gonna see the information. The best way to curb this is to not share your personal information at all.}'' This newfound clarity extended to data access purposes, and participants became more aware of their information being used not only for service improvement but also for training models and other purposes.

\subsubsection{Raise participants’ awareness of privacy risks in LLMs (P1, P2, P3, P7, P9)} 
The use of CLEAR assists participants in understanding and clarifying possible privacy risks when using LLMs. 
Initially, participants only had a general understanding of privacy risks, but the tool provided detailed insights that clarified these risks. 
P1, for example, noted that CLEAR highlighted specific areas where their data was at risk, allowing a better understanding of potential privacy issues. She said ``\textit{The tool showed me exactly the areas where it was at risk and put it into focus so that I could have a better understanding of it. [Initially,] I had a general vague fuzzy idea of what the risk was and then the tools showed me exactly [the risks]...The tool removes the vagueness in my head of what LLM could actually do.}'' 
Similarly, as shown in Table~\ref{tab: comparison}, P4 initially believed that data access was only for service improvement, but later recognized the potential for commercial use of the data and sharing of the data with third-parties, stating, ``\textit{My information could be sold. People could find a lot of my information. My information might end up being saved and we're places where it shouldn't be.}''

\subsubsection{Lead to an intention for behavior change (P4, P6, P9)} 
The increased awareness and understanding fostered by the use of CLEAR led our participants to consider changing their behavior to better protect their privacy. 
P6, for example, concluded that using LLMs necessitates more cautious sharing of personal information: ``\textit{The user should reconsider providing the name, address, and email address with the AI and just use the one-sentence summary that he/she is requesting help with...I need to take steps to make sure that I'm at least not providing over information about myself other than basic things like updating a resume}''. 
P9's statement further highlighted the importance of not sharing personal information publicly, ``\textit{The government authorities will be able to access the data, so I think it is wise to not put your personal information on there.}'' This collective change in behavior reflects the participants' intentions to adopt more privacy-conscious practices as a result of their improved understanding of LLM privacy risks.

\subsection{Participants' Reactions to the Disclosure of Sensitive Information}

\subsubsection{Most participants want to remove sensitive information from prompts or replace it with fake data (P1, P2, P3, P5, P6, P7, P8, P11, P12)} 
Our participants expressed a strong desire to remove sensitive information from their prompts or substitute it with fake data. P1 explicitly stated their preference for removing all personal information: ``\textit{I think I would remove all the all the personal information.}'' Similarly, P6 emphasized the importance of excluding personal details such as name, email address, telephone number, and home address: ``\textit{I just removed like my name my email address, my telephone number. And my home address from the prompt.}'' Additionally, P5 highlighted an alternative approach by suggesting the use of fake contact details: ``\textit{I may like disguise him by providing fake email and fake phone number. I will not provide these details and only give a generic name, email, and phone number.}'' These responses indicate a common practice among our participants to ensure their privacy and security by omitting or obfuscating sensitive data in their prompts. Some participants even suggested that CLEAR could directly provide solutions for them, as P4 mentioned, ``\textit{I would like to know more like not only the information they present but also some solutions that I can choose.}''

However, several participants chose to proceed with sending the sensitive information without any modification. For instance, P13 stated, \textit{It wouldn't stop me from using ChatGPT. I'll send it out.''} They believed that this information could still be obtained by other applications in other ways, regardless of whether it was shared here or not. This opinion was further explained by P5: \textit{Because I already know that they [other online services] have access to this information. These days, your email and your phone number or your mailing address are some of the easiest things to get. So, I'm okay that if they are using so.''}

\subsubsection{Participants expect CLEAR to identify more types of sensitive information and define the scope of ``sensitive'' clearly (P1, P2, P4, P5, P5, P7, P12)} \label{sec: sensitive_scope}
Participants showed a strong desire for CLEAR to identify more modalities and types of sensitive information. P2 expresses a need for the system to identify not only text but also sensitive information in images and PDF files: ``\textit{I want it can not only identify this text but also the things in the image and PDF files.}'' P5 emphasizes the importance of support for identifying sensitive information related to financial transactions and health issues: ``\textit{there should be some support for [identifying] like financial transaction and health-related issues.}'' P1 raised concerns about the system's ability to handle less directly sensitive information, questioning how the system would react to seemingly innocuous details like the names of siblings or past residences: ``\textit{I just wonder with information that isn't so directly sensitive. I wonder how the system would act. Let's say I input, I have three sisters and their names. Would that be a sort of red flag for your system? Or let's say I lived in a city when I was younger. Is that a risk because I am not living there now? So I don't know how the system would react. That's what I mean by less sensitive.}'' These insights reflect participants' interest in a more comprehensive and nuanced approach to identifying and managing sensitive information.
\section{Results of Case Study 2}
% \begin{table*}[!ht]
% \centering
% \begin{tabular}{llrr}
% \toprule
% \textbf{Polarity} & \textbf{Statement} & \textbf{Mean} & \textbf{SD} \\ 
% \midrule
% Positive & I think that I would like to use this system frequently. & 4.00 & 0.91 \\
% Positive & I thought the system was easy to use. & 4.77 & 0.44 \\ 
% Positive & I felt very confident using the system. & 4.77 & 0.44 \\ 
% Positive & I found the various functions in this system were well integrated. & 4.46 & 0.52 \\ 
% Positive & I would imagine that most people would learn to use this system very quickly. & 4.85 & 0.38 \\ 
% Positive & I feel the information provided is useful. & 4.85 & 0.38 \\ 
% Positive & I thought the policy snippets matched the input sensitive information. & 4.54 & 0.52 \\ 
% Positive & I thought the privacy risks matched the input sensitive information. & 4.62 & 0.51 \\ 
% Negative & I thought there was too much inconsistency in this system. & 1.38 & 0.65 \\ 
% Negative & I found the system unnecessarily complex. & 1.38 & 0.51 \\ 
% Negative & I think that I would need the support of a technical person to be able to use this system. & 1.08 & 0.28 \\ 
% Negative & I found the system very cumbersome to use. & 1.23 & 0.44 \\ 
% Negative & I needed to learn a lot of things before I could get going with this system. & 1.15 & 0.38 \\ 
% Negative & I feel the product is annoying. & 1.15 & 0.38 \\ 

% \bottomrule
% \end{tabular}
% \caption{Mean and Standard Deviation of the Statements in the Post-Study Questionnaire}
% \label{tab:sus_results}
% \end{table*}

\begin{figure*}
    \centering
    \includegraphics[width=\textwidth]{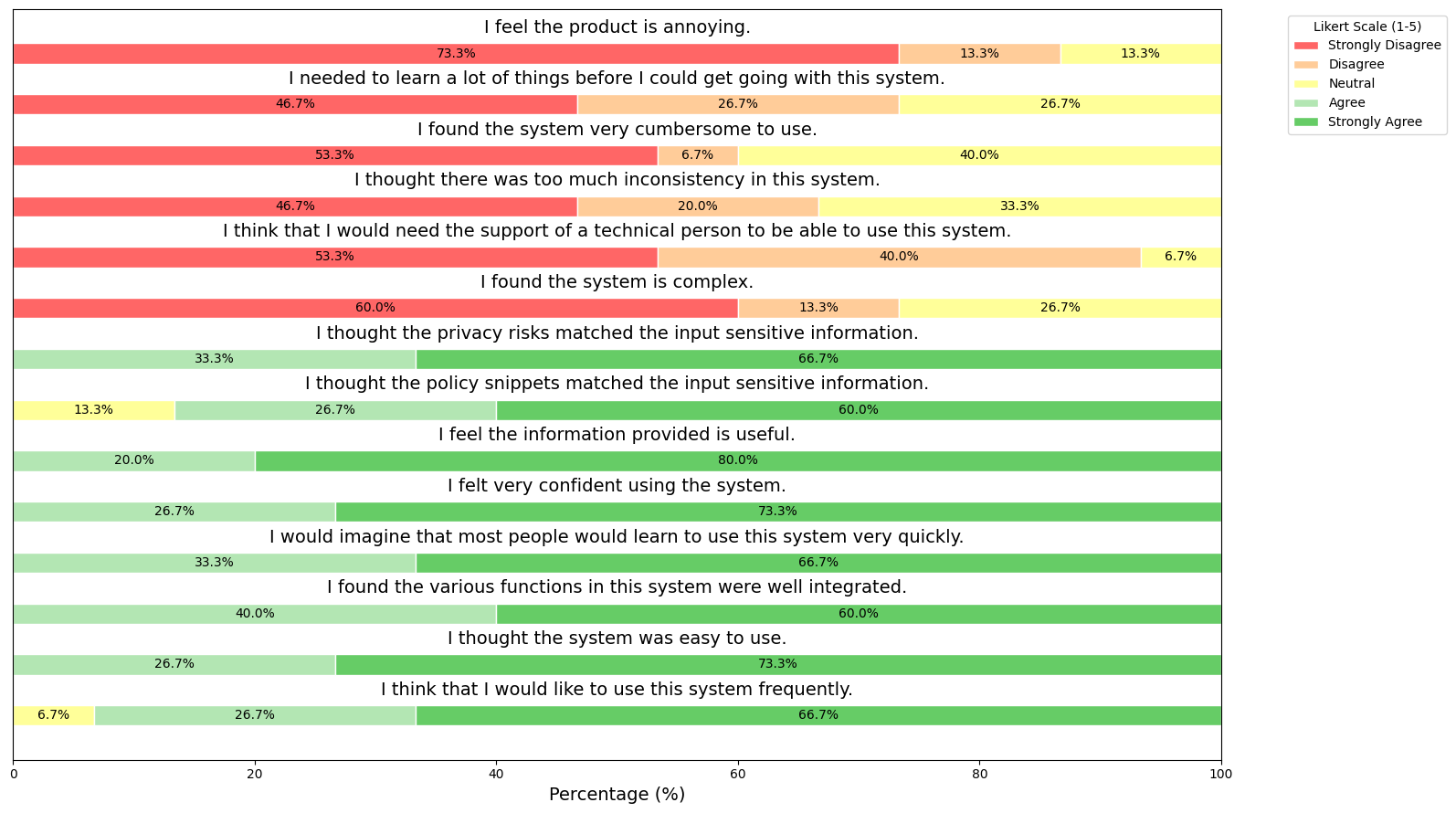}
    \caption{Result of Post-study Questionnaire for the Case Study 2.}
    \label{fig:sus_results_2}
\end{figure*}

\subsection{System Usability}
\label{sec:case study 2 system usabilty}
The evaluation of CLEAR in the context of Gmail used the same assessment method as the first case study using SUS questionnaire~\cite{bangor2009determining}. The results show a positive user experience, as shown in Fig~\ref{fig:sus_results_2}. All participants found the system easy to use, well-integrated, and quick to learn. The confidence in using the system is high, and the information provided is useful. The policy snippets and privacy risks were appropriately matched with sensitive information. 
% Participants do not perceive the system as complex or cumbersome, nor do they feel the need for technical support.

Participants generally found the system useful for improving their awareness of privacy risks. For example, P4 said, ``\textit{So it's actually informing me about the risks. It's saying, `If you do this, here's what happens.' So, in a way, it's guiding me through the process, showing me what I can and cannot include. It helps me understand what to share and what to avoid, making it easier to protect my privacy.}'' P13 also mentioned, ``\textit{This is really good, especially from the user’s perspective, because many people aren’t aware of the risks. They just reply to emails, sometimes including sensitive information like their Social Security number or phone number, without thinking about the consequences. That's why it's important to clearly show what could happen if they share this information.}''

It is worth noting that participants also expressed trust in CLEAR's ability to protect their privacy, noting that although CLEAR is an LLM-based application, it identifies sensitive information locally on the user’s device. Only the types of sensitive information are transmitted to the LLM, not the actual data. As P1 noted, \textit{``[CLEAR] feels more trustworthy because it uses a solution to protect user data by not sharing the exact information with the LLM.''} We refer to this as the ``duality of AI'', that is, AI-powered applications present privacy risks but at the same time, introduce opportunities to address privacy issues more broadly. We will expand on this point in the discussion section. 

\subsection{Impact on Participants' Understandings and Actions Regarding Privacy in LLMs}
Similar to the findings from Case Study 1, Case Study 2 also suggests that using CLEAR increases participants’ awareness of data practices and privacy risks in LLMs and leads to intentions for behavior changes. For example, some participants previously believed that the Gemini plugin processed data fully automatically, but after seeing CLEAR's notifications, they learned that human reviewers might also access their sensitive data. For example, P12 said, \textit{``I did not know there would be any humans reviewing the data. I thought it was just an automated process.''} Other participants also became aware of privacy risks related to LLMs that they were previously unaware of, such as memorization and inference risks.
% As P6 mentioned, \textit{``I didn’t realize it (LLM) had the ability to memorize or store this (sensitive) information... If someone else uses certain prompts to query this information, they might have a chance to access it.''} 
This increased awareness led participants to be more cautious and deliberate in protecting their private data, as we further detail in the following section.

\subsection{Participants' Emotional and Behavioral Reactions to the Disclosure of Sensitive Information}

\subsubsection{Participants felt regretful about their past data-sharing behaviors due to the privacy risks}
Participants (P6, P12, P14) were shocked and surprised by the extent of privacy risks they were previously unaware of. P14 was very surprised to see the risks and emphasized not to share personal information. P6 was not only surprised by these risks but also regretted oversharing client data with AI. She mentioned that she had to discuss this issue with her supervisor: \textit{``Because I've already disclosed way too much information about my clients. Now I am worried about our future clients. So I'm gonna have to have a talk with my supervisor.''} She also suggested an increasing feeling of empowerment from using CLEAR, stating, \textit{``I've been really afraid of AI for a long time, but now I'm feeling hopeful that more things like this are going to come about.''} These strong emotional reactions highlight the positive impact of CLEAR in raising participants’ privacy awareness and encouraging changes in their privacy behaviors.

\subsubsection{Participants become more cautious and deliberate in protecting their private data}
Participants (P2, P3, P4, P5, P8, P9, P10, P11, P12, P14, P15) indicated their increased intention to proactively remove sensitive information from their prompts after seeing CLEAR's notifications. For instance, P2 said \textit{``I would say I have not really paid attention to the privacy concerns in detail. But this (content in the pop-up) is actually enlightening...I think it would prompt me to keep my private information more secure.''} P6 also stated, \textit{``So I would definitely take these (sensitive information) out before asking AI to summarize that (email) for me.''} 

It should be noted that unlike ChatGPT in Case Study 1 which could only access sensitive information from participants' prompts, the Gemini plugin can also retrieve sensitive information directly from emails. Therefore, removing sensitive information from the prompt will not prevent the plugin from accessing it in the emails. Participants mentioned that they would be more cautious and try to avoid sharing sensitive information with the Gemini plugin, or even not using Gemini for emails at all, as P9 mentioned, \textit{``Maybe if I have some information I don't want to be exposed, I might not use the Gemini app.''}. 

Additionally, even though CLEAR focuses on providing privacy awareness in LLM-based applications, participants expressed a desire for the applications to explicitly obtain user consent before accessing their data. As P7 noted, \textit{``It should clearly ask for permission: `Do you allow access to your location?' with options to deny, allow, or decide later. If I decide to allow access to my location, address, IP, or work information, that should be my choice. Ultimately, nothing should happen without our consent.''} To some extent, this desire indicates participants' expectation of more explicit privacy choice, which is largely lacking in today's LLM-based applications~\cite{zhang2024s}. We will further discuss this point in the Discussion section. 
% \yaxing{connect to the literature about notice and choice}

% notice & choice

\section{Discussion}

\subsection{The Duality of AI in Privacy}
In situating our findings within the existing literature on AI and privacy, we identified a notable tension regarding the role of AI in the management of user privacy.  AI exhibits a dual nature in this context.
% AI has a dual nature in the realm of privacy. 
On the one hand, recent studies indicate that AI technologies, such as LLMs, present significant privacy risks.  For example, Lee et al.~\cite{lee2024deepfakes} identified twelve specific risks associated with AI, such as threats in data acquisition, data processing, data dissemination, and facilitation of cyber invasions. These risks underscore AI's potential to compromise user privacy through various channels. Moreover, the lack of transparency surrounding AI operations and its data usage contributes to diverse and often misguided mental models held by users~\cite{zhang2024s}. Despite these concerns, our work revealed the opportunities and user-expressed desires to utilize AI for privacy protection. The evaluation of CLEAR provided empirical evidence supporting the positive impact of an AI-based system on enhancing users' privacy awareness and protective behaviors.

This paradox illustrates the ``duality of AI'' in the realm of privacy. While AI technologies inherently present privacy challenges, they simultaneously offer opportunities to assist users in managing their privacy~\cite{chen2024empathy, kapania2024m}. Interestingly, our findings indicated that 
% may stem from a lack of awareness regarding the full extent of AI's privacy threats. However, 
even when participants were made aware of the potential risks associated with LLMs, they still sought AI's assistance in tasks such as ``\textit{detecting where their information actually goes}'' (G3), ``\textit{scaning whatever it is that I'm typing for, specifically for privacy}'' (G5), and ``\textit{tracking potential risk and take action against it}'' (G1).

This phenomenon may stem from participants' perception of AI as a powerful, personalized privacy protection tool capable of delivering adaptive and contextual information based on their activities. This perspective aligns with the concept of a ``personalized privacy assistant (PPA)'' in the context of smartphones~\cite{almuhimedi2015your, liu2016follow} and the Internet of Things~\cite{das2018personalized, das2018personalized}. However, unlike traditional PPAs that provide personalized recommendations based on privacy profiles, such as smartphone permissions~\cite{liu2016follow}, an AI-based system like CLEAR offers more comprehensive insights into data practices, privacy risks, and the potential consequences of users' ongoing tasks. By focusing on tasks rather than personal profiles, CLEAR adopts a less intrusive approach, reinforcing AI's role in supporting users' privacy management.

CLEAR itself capitalizes on the duality of AI. It operates under the assumption that typical LLMs, such as ChatGPT, lack transparency regarding their data practices. Users often have incomplete or varied mental models of how LLMs function~\cite{zhang2024s}, leading to diverse privacy concerns. The current iteration of CLEAR acts as a transparency tool, informing users about hidden data practices, associated privacy risks, and potential consequences. Our study results in Section~\ref{sec:case study 2 system usabilty} show that users trust CLEAR because it is more transparent and less invasive of their privacy when using LLMs. This duality allows CLEAR to effectively utilize AI to enhance transparency in AI systems.

This distinctive characteristic of AI within the privacy domain provides CLEAR with significant advantages over other approaches documented in the literature.  For example, the IoT privacy nutrition labels provide data practice information to users in an easy-to-understand format~\cite{emami2021informative} but would require users to proactively check the labels before purchasing the devices. Privacy Q\&A uses NLP techniques to extract information from privacy policies, allowing users to ask questions about their privacy~\cite {ravichander2019question}. Yet, it does not provide users with the ability to associate the extracted information with the contexts of their tasks. Leveraging the duality of AI, CLEAR not only overcomes these hurdles but also leaves room to expand itself to other task domains (e.g., customer service, online medical consultant, etc.) and platforms (e.g., mobile devices, the Internet of Things).

\subsection{Design Implications}
Based on our findings, we propose design implications for the development of future techniques and policies aimed at enhancing users' privacy awareness related to LLM-enabled end-user applications. While in this work, we present CLEAR as an add-on to existing LLM-enabled end-user applications; we hope that their designers can consider incorporating them into the applications themselves.

\subsubsection{Integrate privacy notification seamlessly with the user experience by providing contextual and real-time privacy information} The previous study~\cite{bergmann2008testing} has demonstrated that users' privacy awareness increases significantly when privacy information is presented within the relevant context. Our study confirms this, revealing that users expect to be informed of potential privacy risks and their implications to the current task context without disruptions to their workflow.  We suggest that future LLM-based systems should provide users with contextual privacy information to increase system transparency and help users make informed decisions. For instance, when users input sensitive personal information into a conversational agent, a brief, context-specific description of potential privacy risks should be displayed directly within the chat interface. Future work should also consider integrating this feature natively in conversational agents and aligning it with its specific data practices to increase its accuracy, precision, and user trust. 
% include contextual information For example, one of the participants in G3 said, ``\textit{I expect it shows two or three lines about the risk related to the content users have written.}'' Similarly, participants in G5 also expressed that ``\textit{[I hope] it can provide like five main points, such as `Are you aware that you are giving permission for XYZ?' so that you can quickly understand what you're agreeing to at a glance.}'' 

\subsubsection{Offer users fine-grained controls over privacy information contents} 

Participants expressed varying preferences regarding the level of detail they want in privacy information. Some preferred concise, essential information at a glance, while others expected access to detailed explanations and full privacy policies. We recommend that future tools allow users to customize the amount of information they receive. A layered approach to presenting information, where users can define the depth of detail they want, could be especially effective in addressing individual needs and constraints.

%Participants have different expectations regarding the level of detail in privacy information. Some expected to access concise, essential privacy details at a glance, while others wanted to delve deeper into the full privacy policy and detailed risk explanations as per their preference. We suggest that future tools should allow users to customize the information they receive. 
% For instance, a participant of G5 mentioned that ``\textit{I can hurry up and read it quickly ... but for those who want to take the time and ... if they want to, they can click on a section [to review more details].}''
% For example, a layered approach to presenting information to users can be promising, allowing users to self-define the levels of information based on their needs and other constraints.

\subsubsection{Allow users to define the scope of sensitive data} Different users have varying interpretations of sensitive information. As discussed in Section~\ref{sec: sensitive_scope}, participants were unclear about whether certain types of ``weak'' sensitive information (e.g., others' personal information, past information, etc.) could be shared with LLMs. Furthermore, what is considered private often depends on the context and individual values. Therefore, it is important not only to inform users about which data is sensitive but also to give them the ability to specify the types of information they are particularly concerned about.

% Additionally, whether the information is considered private often depends on the context and personal value judgments. Therefore, it is necessary not only to alert users about which information is sensitive but also to provide them with the ability to define the types of information they are particularly concerned about.

\subsubsection{Incorporate explicit user consent mechanisms}
While CLEAR focuses on increasing users' privacy awareness, participants expressed the need for explicit consent before LLMs access sensitive data. Existing LLM-based applications only obtain users' consent upon signing up for the services without the ability to change their consent during the interaction. This may become a critical issue as users' content may change when they become more aware of the potential privacy risks using systems like CLEAR. We recommend integrating these consent features into LLMs to provide fine-grained, transparent, and user-driven control, encouraging active data-sharing decisions and building trust in LLM-based applications.

\subsection{Limitation and Future Work}
We summarize the limitations of our work in both the study design and the system design, and we propose future steps to address these limitations.

% in Section 8.3.1, we will discuss the future plan to expand the study scale and conduct a longitudinal deployment study. The future study will recruit a larger and more representative participant group through online recruitment platforms (e.g., Prolific), ensuring diversity across demographics, levels of technical proficiency, and privacy awareness backgrounds. Participants will install the CLEAR extension and interact with LLM conversational interfaces or LLM-based add-ons in the real context of use for 2 weeks. This future longitudinal study will allow us to evaluate the ecological validity of CLEAR and the impact of its use on the user’s real privacy behaviors. 

\subsubsection{Study design}
Although the results of the evaluation study suggest the high usability and usefulness of our system, they are limited to a short-term study in a lab setting. A future \chaoran{longitudinal} deployment study is needed to validate these findings in long-term and real-world settings. Our study is an initial effort to enable users to investigate relevant privacy policies and potential privacy risks when using LLMs. In the future, \chaoran{we will recruit a larger and more representative participant group through online recruitment platforms (e.g., Prolific), ensuring diversity across demographics, levels of technical proficiency, and privacy awareness backgrounds. Participants will install the CLEAR extension and interact with LLM conversational interfaces or LLM-based add-ons in the real context of use for 2 weeks. We plan to conduct pre-tests and post-tests with users to assess their acquisition of privacy knowledge and collect additional behavioral data when users interact with CLEAR in the deployment (similar to \cite{story2021awareness, story2022increasing}).} 
% This will help identify the key factors and issues that influence users' privacy awareness when using our system. 
\chaoran{This future longitudinal study will allow us to evaluate the ecological validity of CLEAR and the impact of its use on the user’s real privacy behaviors. }

\subsubsection{System design}
Our current CLEAR prototype serves primarily as a proof of concept to facilitate the experiment discussed in this paper.  At present, it is compatible only with ChatGPT and the Gemini plugin for Gmail in order to demonstrate the feasibility of CLEAR in two primary use cases (standalone LLM conversational interfaces and the LLM add-ons for GUI applications). While ChatGPT and Gmail are the most popular applications in their category, with ChatGPT having a 60\% market share~\cite{westfall2023chatgpt} and Gmail holding a 53\% share of the US email market\footnote{\url{https://techreport.com/statistics/software-web/gmail-statistics/}}, there are other popular LLMs (e.g., Claude) and applications (e.g., Microsoft Copilot) that the current CLEAR tool does not support. 
\chaoran{However, the key features of CLEAR do not rely on any specific technicalities of ChatGPT or Gemini for Gmail, and it is straightforward to extend CLEAR to support other chat-based LLM interfaces (e.g., Claude, web UIs for LLaMA and Qwen) and web-based LLM add-ons. 
To expand its scalability, we open-sourced CLEAR~\footnote{https://github.com/CRChenND/CLEAR}. CLEAR's extensible architecture will make it easier to integrate CLEAR with other popular LLMs’ GUI in the future. }

Our system also has limited capabilities in identifying additional modalities of sensitive information (e.g., through voice interaction) and handling sensitive information contained in files such as images and PDFs uploaded to LLMs. Moreover, while we effectively use the Microsoft Presidio library to detect emails, contacts, and locations, CLEAR currently lacks the capability to promptly identify other types of personally identifiable information, such as financial and health data. \chaoran{To enhance CLEAR's functionality, we plan to investigate using small-scale, open-source language models to identify sensitive information on the edge. Recent advancements~\cite{zhou2024rescriber, li2024governing} allow these models to run locally on the client side, allowing for more accurate detection of sensitive information across a broader range of types, while keeping user data on-device and mitigating privacy risks.}

\chaoran{Currently, CLEAR uses text to summarize privacy policies and generate privacy risks as an initial effort to enable users to investigate relevant privacy policies and potential risks when using LLMs. For users with limited privacy literacy or technical expertise, textual descriptions may not be the most efficient way to communicate the implications of privacy practices or the significance of identified risks.
With the development of multimodal LLMs, there is an opportunity to overcome these limitations by incorporating more interactive and accessible explanations. Future work could focus on integrating multimodal features into CLEAR, such as interactive tutorials or visual summaries that highlight key aspects of privacy policies and risks. For example, users could explore how their data might be processed or shared through animations or flowcharts that depict data flows. Integrating multimodal LLMs into CLEAR has the potential to transform how privacy policies and risks are communicated, making them more transparent, engaging, and comprehensible for a wider audience.}

\section{Conclusion}
In response to the emerging privacy risks posed by the use of LLMs in end-user applications, we conducted five co-design workshops to study user needs, preferences, and constraints relevant to making informed privacy decisions when interacting with LLM-enabled end-user applications. Based on our findings, we developed CLEAR, a tool aimed at enhancing user understanding of privacy policies and associated risks in LLM-enabled applications. We conducted two user studies, one with the representative standalone LLM application ChatGPT and another with the Gemini ad-don for Gemail, demonstrating that CLEAR significantly improves privacy awareness and encourages safer data-sharing practices. The insights derived from our research offer important design implications for the development of AI-based tools that empower users to manage their privacy more effectively.

\begin{acks}
This work was supported in part by a Google Research Scholar Award, a Notre Dame-IBM Technology Ethics Lab Award, a Google PSS Privacy Research Award, NSF Grant CNS-2426395, NSF Grant CNS-2426397, and NSF Grant CNS-2232653. C. Chen and Y. Ye’s work was partially supported by the NSF under grants IIS-2321504, IIS-2334193, IIS-2217239, CNS-2426514, CNS-2203261, and CMMI-2146076. Any opinions, findings, and conclusions or recommendations expressed in this material are those of the authors and do not necessarily reflect the views of the sponsors. We would like to thank Allen Yilun Lin for useful discussions.
\end{acks}

\bibliographystyle{ACM-Reference-Format}
\bibliography{Mybib}

\newpage
\appendix
\section*{A. Codebook for the participatory design \textsuperscript{1}} 
\footnotetext[1]{The numbers in parentheses indicate the number of participants associated with each code in the codebook.}
\label{codebook1}

% \documentclass{article}
% \usepackage{enumitem} % Load package for list customization

% \begin{document}

% \section*{Study 1 Codes}
\begin{enumerate} 
    \item {User needs}
    \begin{enumerate}
        \item Improve privacy Literacy 
        \begin{enumerate}[label=(\roman*)] % Roman numerals for nested items
            \item Force users to spend more time understanding policies (3)
            \item Show connection between data usage and system features (3)
            \item Have a tutorial to go through all security features in the system (2)
            \item Raise users' awareness by showing the value of their data (2)
            \item Simulating the scams to help people learn to react to them (2)
            \item Have courses to improve people's privacy literacy (1)
        \end{enumerate}
        \item Use AI to protect privacy
         \begin{enumerate}[label=(\roman*)] % Roman numerals for nested items
            \item Inspect sensitive info in user input (4)
            \item Have LLMs unlearn data (1)
            \item Use AI to monitor users and defend against hacking (10)
            \item Detect data traffic and block risky ones (1)
        \end{enumerate}
        \item Enable users to take action
          \begin{enumerate}[label=(\roman*)] % Roman numerals for nested items
            \item Help users do data sanitation (1)
            \item Help users recognize and manage the data they share (10)
            \item Provide users with an option to either skip reading it or read it (2)
            \item Give users options to accept, deny, or modify info to submit (4)
        \end{enumerate}
        \item Information presentation
         \begin{enumerate}[label=(\roman*)] % Roman numerals for nested items
            \item Show reasons and consequences of privacy data usage
            \begin{enumerate}[label=\Alph*.] % Roman numerals for nested items
            \item Use animation to show the consequences of giving consent (2)
            \item Show consequences of sharing certain private data (5)
        \end{enumerate}
            \item Information should be adjusted to user groups
            \begin{enumerate}[label=\Alph*.] % Roman numerals for nested items
            \item Explain the privacy segments for people in every generation (1)
            \item The tool should be accessible to all user groups (7)
        \end{enumerate}
            \item Information should be compact and easy to understand
             \begin{enumerate}[label=\Alph*.] % Roman numerals for nested items
            \item Add tooltips to show definitions of words in policy (1)
            \item Visualize important info on one page and show others in links (5)
            \item Add link to the detailed information (6)
            \item Use simple language to summarize the information (8)
            \item Show information concisely (3)
        \end{enumerate}
        \end{enumerate}
        \item Contextual privacy policy and risk alert
         \begin{enumerate}[label=\Alph*.] % Roman numerals for nested items
            \item Summarize the share of private data and potential privacy risks (1)
            \item Indicate users of sensitive information and possible reactions (5)
            \item Notify relevant policy segments and consequences (6)
            \item Alert themselves if a privacy breach happens (8)
            \item Show brief and in-context privacy notice (2)
        \end{enumerate}
        \end{enumerate}
    \item{Existing privacy protection methods}
    \begin{enumerate}
        \item Self-data sanitation
        \begin{enumerate}[label=(\roman*)] % Roman numerals for nested items
            \item Disguise real location and online behavior data with fake ones (1)
            \item Use fake name online (2)
            \item Anonymize data and disassociate data (3)
        \end{enumerate}
        \item Self-privacy education
        \begin{enumerate}[label=(\roman*)] % Roman numerals for nested items
            \item Check app permission regularly (2)
            \item Educate themselves on privacy risks (5)
        \end{enumerate}
        \item Secure browsing practices
         \begin{enumerate}[label=(\roman*)] % Roman numerals for nested items
            \item Use different usernames/passwords for every site (8)
            \item More strict authentication (4)
            \item Use a more secure browser to search (1)
        \end{enumerate}
        \item Limit data sharing
         \begin{enumerate}[label=(\roman*)] % Roman numeral
          \item Have control over the data being shared (8)
          \item Store privacy data locally (1)
          \item Not sharing sensitive information to risky app or people (11)
          \item Limit the data types that users can upload to the system (2)
    \end{enumerate}
    \end{enumerate}

    \item {Things in privacy policies that people care about}
    \begin{enumerate}
        \item Show detailed information about who will use the privacy data (6)
        \item How the data access influences users
         \begin{enumerate}[label=(\roman*)] % Roman numeral
          \item Influence of privacy policy updates (2)
          \item Security and privacy risks (2)
    \end{enumerate}
        \item What data is being accessed
        \begin{enumerate}[label=(\roman*)] % Roman numeral
          \item What kind of data the app will access (4)
          \item Payment and transaction (1)
          \item Access to contact information (1)
          \item Frequently used service (1)
          \item Tracking of online behavior (2)
    \end{enumerate}
        \item Why is data being accessed (3)
    \end{enumerate}

    \item {Reasons for not reading privacy policies}
    \begin{enumerate}
        \item Privacy resignation
        \begin{enumerate}[label=(\roman*)]
        % Roman numeral
          \item Users think service providers don't want them to read (1)
          \item Users think their data will always be leaked (8)
          \item Users have to use the service (2)
          \item Users think all privacy policies are the same (5)
    \end{enumerate}
        \item Privacy policy is difficult to understand
        \begin{enumerate}[label=(\roman*)]
        % Roman numeral
          \item Design issue/not user friendly (2)
          \item Difficult to understand (6)
          \item Users feel uncertain to identify (2)
    \end{enumerate}
        \item Privacy policy is too long
          \begin{enumerate}[label=(\roman*)]
        % Roman numeral
          \item Privacy policies are too long (10)
          \item Users don't want to spend a lot of time reading and understanding (2)
    \end{enumerate}
    \end{enumerate}
\end{enumerate}
\section*{B. Structured Privacy Knowledge} \label{appendix: structured privacy knowledge}

\begin{enumerate}
    \item \textbf{Privacy Risk:} Data breaches and sale of personal data
    \begin{itemize}
        \item \textbf{Cause:} LLM-based CAs operating on the cloud
        \item \textbf{Influence:} Users losing control over their chat logs
        \item \textbf{Evidence:} Most popular LLM-based CAs operate on the cloud
    \end{itemize}
    \item \textbf{Privacy Risk:} Memorization risks
    \begin{itemize}
        \item \textbf{Cause:} LLMs memorizing details in the training data
        \item \textbf{Influence:} Sensitive information being leaked in response to prompts
        \item \textbf{Evidence:} LLMs using user data to train their models periodically
    \end{itemize}
    \item \textbf{Privacy Risk:} Disclosure of personally identifiable information (PII)
    \begin{itemize}
        \item \textbf{Cause:} Users disclosing their own and others' data in conversations
        \item \textbf{Influence:} Implicating interdependent privacy issues
        \item \textbf{Evidence:} Users disclosing various types of PII in chat histories
    \end{itemize}
    \item \textbf{Privacy Risk:} Dark patterns in opt-out interfaces
    \begin{itemize}
        \item \textbf{Cause:} Discouraging users from exercising privacy controls
        \item \textbf{Influence:} Users feeling they have to sacrifice privacy for benefits
        \item \textbf{Evidence:} Opt-out interfaces linking privacy and utility loss
    \end{itemize}
    \item \textbf{Privacy Risk:} Extraction of personal attributes from text
    \begin{itemize}
        \item \textbf{Cause:} LLMs lack commonsense about social privacy norms
        \item \textbf{Influence:} Malicious actors can infer personal attributes from seemingly harmless text
        \item \textbf{Evidence:} Inference of location based on text mentioning a specific traffic maneuver
    \end{itemize}
    \item \textbf{Privacy Risk:} Lack of understanding of privacy norms
    \begin{itemize}
        \item \textbf{Cause:} LLMs lack commonsense about social privacy norms
        \item \textbf{Influence:} Difficulty in keeping secrets and protecting privacy
        \item \textbf{Evidence:} Models can be easily tricked by third-party adversaries to ignore privacy-protecting instructions
    \end{itemize}
    \item \textbf{Privacy Risk:} User sharing sensitive information with LLM-based CAs
    \begin{itemize}
        \item \textbf{Cause:} High utility and human-like interactions of LLM-based CAs
        \item \textbf{Influence:} Users sharing sensitive and personally identifiable information
        \item \textbf{Evidence:} Users constantly facing challenges in protecting their privacy due to flawed mental models and dark patterns in privacy management features
    \end{itemize}
    \item \textbf{Privacy Risk:} Exposure risk
    \begin{itemize}
        \item \textbf{Cause:} AI technologies can generate human-like media, such as deepfake pornography, without consent
        \item \textbf{Influence:} AI technologies can lead to the unauthorized dissemination of sensitive or private information
        \item \textbf{Evidence:} Twitch streamer QTCinderella's plea to stop spreading links to AI-generated deepfake pornography
    \end{itemize}
    \item \textbf{Privacy Risk:} Phrenology/physiognomy risk
    \begin{itemize}
        \item \textbf{Cause:} AI can learn arbitrary classification functions and potentially infer sensitive attributes like sexual orientation from physical features
        \item \textbf{Influence:} AI technologies can perpetuate harmful stereotypes and discrimination based on physical appearance
        \item \textbf{Evidence:} The belief that AI can automatically detect things like sexual orientation from physical attributes
    \end{itemize}
    \item \textbf{Privacy Risk:} Surveillance risk
    \begin{itemize}
        \item \textbf{Cause:} Facial recognition classifiers require large amounts of face data, leading to uncritical data collection practices
        \item \textbf{Influence:} AI technologies can enable widespread surveillance and tracking of individuals without their consent
        \item \textbf{Evidence:} Collection of face scans in airports for facial recognition purposes
    \end{itemize}
\end{enumerate}
% \chaoran{Include the number of participants associated with each code in the codebook to improve the transparency and representativeness of key insights.}

\section*{C. Codebooks \textsuperscript{1}} 
\footnotetext[1]{The numbers in parentheses indicate the number of participants associated with each code in the codebook.}

\label{codebook2}

\subsection*{Case Study 1}
\begin{enumerate} 
    \item {User feedback}
     \begin{enumerate}
      \item Identify the access to specific services, such as text messaging or conversations (1)
      \item Provide users options to look into more details (1)
      \item Show potential solutions to users (2)
      \item Identify more types of sensitive info (image/PDF/financial/health) (4)
      \item Why the number of policy snippets differ (1)
      \item How to definite the scope of sensitive information (3)
      
      \end{enumerate}
    \item {Changes brought yo users after using the tool}
     \begin{enumerate}
     \item Compare benefit and risk, if the risk is not bad, accept it (5)
     \item Intention for behavior change (2)
     \item Raise more privacy concerns (2)
     \item Be aware of new reasons for data accessing (2)
     \item Be aware of new privacy risks (4)
     \item Clarify privacy risks (3)
     \item Be aware of new data accessors (11)
     \end{enumerate}
    \item {Reaction to the prompt after seeing the tool}
      \begin{enumerate}
      \item Use fake/dummy info to replace sensitive info (7)
      \item Use placeholder to replace sensitive info (1)
      \item Remove sensitive info (5)
     \end{enumerate}
\end{enumerate}

\subsection*{Case Study 2}
\begin{enumerate} 
    \item {Participants' emotion}
     \begin{enumerate}
      \item Feels hopeful and excited about the extension (1)
      \item Be shocked and surprised about the privacy risks (5)
      \item Regrets oversharing information with AI (1)
      \end{enumerate}
    \item {Behavior change}
     \begin{enumerate}
     \item Refuse to use LLM if any sensitive information is involved (1)
     \item Remain no change due to privacy resignation (2)
     \item Desire to keep certain sensitive information private (4)
     \item Actively removing personal data from their prompts (8)
     \item Be more cautious and deliberate in protecting their privacy data (14)
     \end{enumerate}
    \item {Participant's suggestions}
      \begin{enumerate}
      \item Shows consequences of privacy data leak(2)
      \item The information can be more concise (4)
      \item Identify sensitive information in the attachments(1)
      \item Ask for the client's consent about sharing the sensitive info (1)
      \item Useful to know before sending an email out (1)
      \item The extension should allow users to not share personal info (1)
      \item Websites should get user consent before accessing personal data (1)
      \item Participants want to know more about how data may be leaked (3)
      \item Add 2FA or passkeys before sharing personal info (1)
     \end{enumerate}
      \item {Comment on usability and usefulness}
      \begin{enumerate}
      \item Useful to highlight sensitive info, especially for long emails (1)
      \item Help verify privacy hypothesis (1)
      \item Plans to inform their supervisor about the data leakage (1)
      \item Highly impressed and appreciative of the detailed information (1)
      \item The extension is easy to use and the function is positive (15)
      \item The extension is trustful since it protects users' exact data (2)
      \item The extension is explanatory and easy to understand (8)
      \end{enumerate}
      \item {Improve participants' privacy awareness}
      \begin{enumerate}
      \item Improve users' awareness about privacy risks (2)
      \item Sensitive info may be collected without explicit consent (2)
      \item Understand flawed mental model and dark pattern (1)
      \item Sensitive information can be used for controlled access (2)
      \item LLMs may lack conformity to social privacy norms (2)
      \item Someone can prompt and LLM to reveal personal info (3)
      \item Memorization risk (10)
      \item The reason why Google want to access phone number or other information (2)
      \item Human reviewers might access sensitive data (7)
      \end{enumerate}
\end{enumerate}

\end{document}